\documentclass[useAMS,usenatbib]{mn2e}

\usepackage{subeqnarray,graphicx,amsmath}
\usepackage{multirow}
\newcommand{\MS}{\ifmmode{\,}\else\thinspace\fi{\rm M}\ifmmode_{\odot}\else$_{\odot}$\fi}
\newcommand{\LS}{\ifmmode{\,}\else\thinspace\fi{\rm L}\ifmmode_{\odot}\else$_{\odot}$\fi}
\newcommand{\RS}{\ifmmode{\,}\else\thinspace\fi{\rm R}\ifmmode_{\odot}\else$_{\odot}$\fi}
\title[Chaos in hydrodynamic BL~Herculis models]{Chaos in hydrodynamic BL~Herculis models.}
\author[R. Smolec \& P. Moskalik]
{R. Smolec\thanks{E-mail:
smolec@camk.edu.pl} and
P. Moskalik\\
N. Copernicus Astronomical Centre, Bartycka 18, 00-716 Warszawa, Poland\\
}

\begin{document}

\date{Accepted . Received ; in original form }

\pagerange{\pageref{firstpage}--\pageref{lastpage}} \pubyear{2011}

\maketitle

\label{firstpage}

\begin{abstract}
We present non-linear, convective, BL~Her-type hydrodynamic models that show complex variability characteristic for deterministic chaos. The bifurcation diagram reveals a rich structure, with many phenomena detected for the first time in hydrodynamic models of pulsating stars. The phenomena include not only period doubling cascades en route to chaos (detected in earlier studies) but also periodic windows within chaotic band, type-I and type-III intermittent behaviour, interior crisis bifurcation and others. Such phenomena are known in many textbook chaotic systems, from the simplest discrete logistic map, to more complex systems like Lorenz equations.

We discuss the physical relevance of our models. Although except of period doubling such phenomena were not detected in any BL~Her star, chaotic variability was claimed in several higher luminosity siblings of BL~Her stars -- RV~Tau variables, and also in longer-period, luminous irregular pulsators. Our models may help to understand these poorly studied stars. Particularly interesting are periodic windows which are intrinsic property of chaotic systems and are not necessarily caused by resonances between pulsation modes, as sometimes claimed in the literature. 
\end{abstract}


%%%%%%%%%%%%%%%%%%%%%%%%%%%%%%%%%%%%%%%
\section{Introduction}\label{sec.intro}
%%%%%%%%%%%%%%%%%%%%%%%%%%%%%%%%%%%%%%%

Chaotic dynamics is present in many astrophysical systems and stellar variability is not an exception, although in this case, chaos was studied mostly in the context of hydrodynamic models of large amplitude pulsators. \cite{bkov87} and \cite{kovb88} found a chaotic behaviour in their radiative type-II Cepheid models (W~Vir and RV~Tau). In depth analysis of chaos in these models was conducted by \cite{skb96} and \cite{let96}. Also \cite{bm92} found chaotic behaviour in two sequences of radiative BL~Her-type models, however did not analyse the phenomenon. Recently chaotic behaviour was reported in convective hydrodynamic models of BL~Her stars \citep{sm12} and RR~Lyrae stars \citep{pkm13}.

On observational side chaos was detected in type II Cepheids of RV~Tau type \citep[R~Scuti and AC~Her;][]{bks96,kbsm98} and in several semi-regular variables \citep{bkc04}, and in Mira-type variable \citep{ks03}.

In this paper we report on the chaotic behaviour we have found in a sequence of non-linear convective BL~Her models. For the first time in stellar pulsation modelling we clearly demonstrate the appearance of dynamical phenomena  well known and common to classical chaotic systems, both discrete (e.g. logistic or H\'enon maps) and continuous (e.g. R\"ossler or Lorenz equations). In all these systems the basic route to chaos is through a cascade of period doubling bifurcations also present in our models and in earlier studies of radiative type-II Cepheid models \citep{kovb88}. In addition, our models display a full wealth of dynamic behaviour characteristic for deterministic chaos. Within chaotic band we find several windows of non-chaotic variation (windows of order), with stable period-$n$ limit cycles. These windows are either extremely narrow or relatively large. In the latter case the periodic window is preceded by the type-I intermittent behaviour, till the periodic cycle is born through the tangent bifurcation. This periodic cycle again undergoes a series of period doubling bifurcations en route to chaos. The interior crisis bifurcations, in which separate chaotic bands merge, leading to the abrupt increase of the attractors volume are detected in our models, as well as crisis induced intermittency and type-III intermittency.

Chaos was not detected in any BL~Her star so far. In our opinion however, these models are important for several reasons. ({\it i}) Chaos does occur in larger-luminosity type-II Cepheids -- RV~Tau stars, as well as in semi-regular variables. Our models may shed more light on variability of these poorly studied stars. ({\it ii}) We initiated the survey of non-linear convective pulsation models of type-II Cepheids extending to the highest luminosities (RV~Tau domain) in which chaotic variability is expected, as previous radiative models and observations indicate. In the present paper we introduce and test the methods to study chaos in such models. ({\it iii}) The striking similarity between our hydrodynamic models of pulsating stars and even the simplest chaotic systems, like logistic map, is noteworthy, indicating that many very different systems may share the same dynamical properties. ({\it iv}) Finally, although the chaotic behaviour was not detected in any BL~Her star so far, we cannot exclude such possibility in the future. We note that the period doubling effect in these stars was predicted by \cite{bm92}, based on radiative hydrodynamic models, but it took 20 years to discover the effect in the first star of this type \citep{igor11,ssm12}.

In Section~\ref{sec.logistic} we summarize the properties of the logistic map, which will help us better understand phenomena occurring in our hydrodynamic models. The reader familiar with the chaos theory may safely skip this Section. In Section~\ref{sec.hydro} we briefly describe the computation and basic properties of the models. In the next Sections we present detailed analysis of the models showing both chaotic and periodic variation, including discussion of the largest Lyapunov exponents (Section~\ref{sec.lyap}). We discuss our results in Section~\ref{sec.concl} and comment on observability of the chaotic phenomena in Section~\ref{sec.obs}. Summary in Section~\ref{sec.summary} close the paper.

Initial results of this study were reported in the conference proceedings of IAU Symposium No 301 ({\it Precision Asteroseismology}), \cite{sm13}.

%%%%%%%%%%%%%%%%%%%%%%%%%%%%%%%%%%%%%%%%%%%%%%
\section{Logistic map}\label{sec.logistic}
%%%%%%%%%%%%%%%%%%%%%%%%%%%%%%%%%%%%%%%%%%%%%%

\begin{figure*}
\centering
\resizebox{\hsize}{!}{\includegraphics{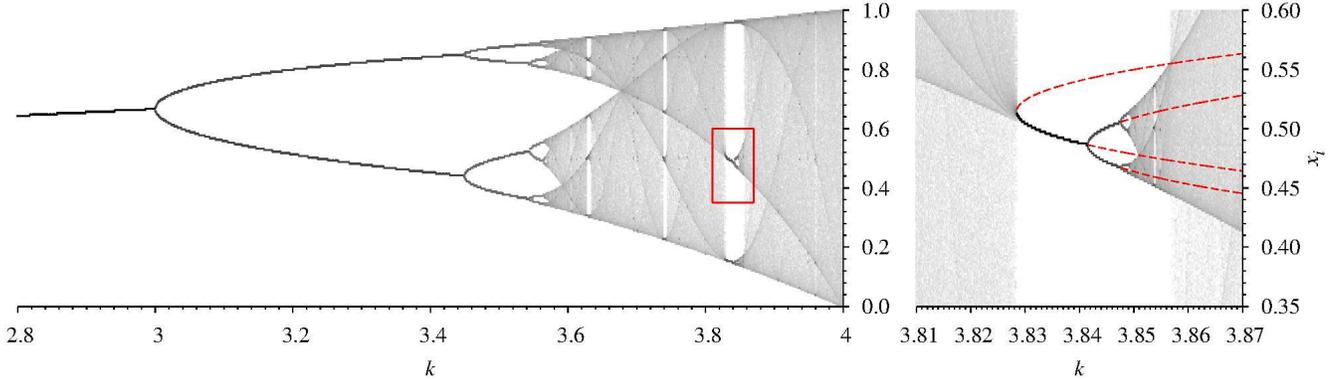}}
\caption{Bifurcation diagram for the logistic map. In the right panel we plot the zoom of the area marked with box in the left panel. Some of the unstable period-3 and period-6 cycles are marked with dashed lines. {\it[This is low resolution version of the Figure]}}
\label{fig.log_bifurcation}
\end{figure*}

In this Section we briefly discuss the properties of the logistic map, which is the simplest 1D discrete system showing deterministic chaos. The map is defined as:
\begin{equation}x_i=f(x_{i-1})=kx_{i-1}(1-x_{i-1})\,,\label{eq.logistic}\end{equation}
where $k$ is a parameter. We are interested in a range $1<k\le 4$ as then the iterates of eq.~\eqref{eq.logistic} are bounded, $x_i\in[0\!:\!1]$. For $k<1$ iterations converge to 0 and for $k>4$ they diverge. The great advantage of the logistic map is its simplicity -- most of the properties, e.g. bifurcation points, fixed points and their stability, may be computed analytically. Full understanding of the mechanisms behind the observed behaviours is thus possible. At the same time logistic map displays nearly all types of behaviours characteristic for deterministic chaos in more complex, continuous and higher dimension systems, that are also present in our hydrodynamic models. In the case of our models analytical approach is  not possible and we are left with the complex output of pulsation code. Comparison with results discussed in this Section allows for a better understanding of our hydrodynamical models. The properties of the logistic map described below may be found in numerous textbooks \citep[e.g.][]{book.PJS} and original articles \citep[e.g.][]{may}, and are given here without derivation.

Depending on the value of $k$ the iterations of any initial $x_0$ (trajectory) either converge to a periodic cycle, or not, and then the chaotic attractor is present. In Fig.~\ref{fig.log_bifurcation} we show the bifurcation diagram -- a possible long-term values of $x_i$ (initial iterations omitted) as a function of $k$. The plot is a stack of grey-scaled histograms: for each value of $k$, we computed several thousand iterations of eq.~\eqref{eq.logistic} and calculated the probability with which the iterations fall to one of the 200 bins into which the $[0\!:\!1]$ interval was divided. Grey bands of chaos are clearly visible, as well as windows of order in which stable periodic cycles are present. The right part of the figure shows the zoom of the largest, period-3 cycle window.

To visualise the attractor it is useful to construct the first return maps, i.e. plots of $x_{i+i}$ vs. $x_i$ (omitting the initial transient). For the discussion below it is instructive to consider the evolution of return map as $k$ is increased. In a range $2.8\le k\le 4$ it is shown in the animation that may be found in the online version of this article as additional supporting information.

\begin{figure*}
\centering
\resizebox{\hsize}{!}{\includegraphics{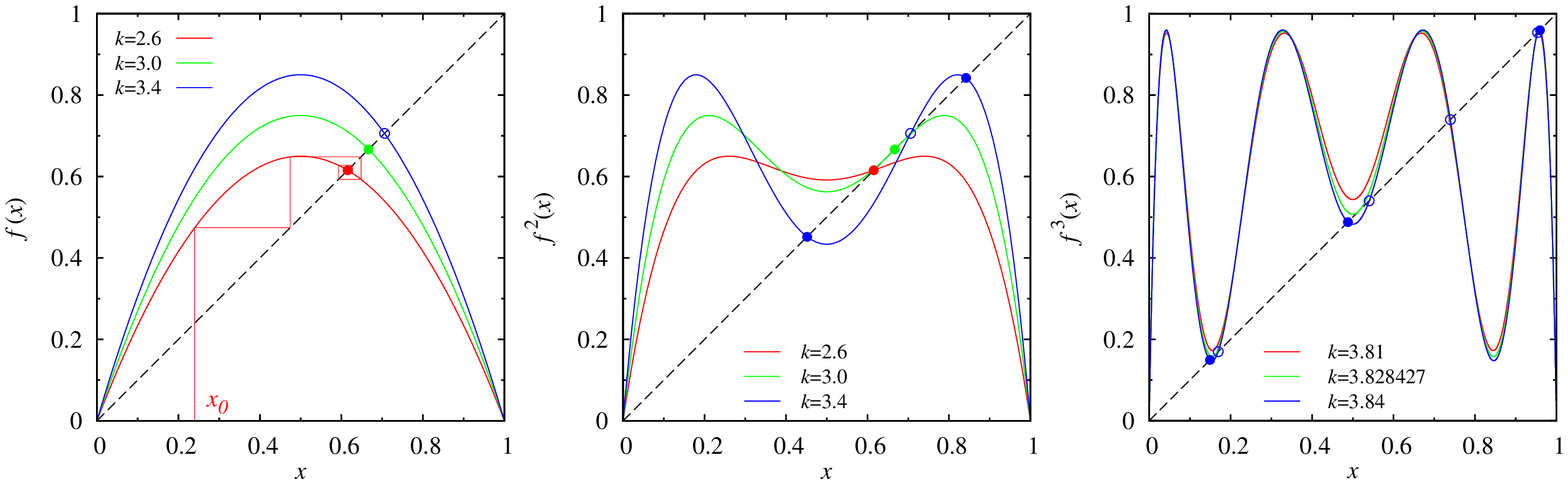}}
\caption{Graphical interpretation of the logistic map. Consecutive iterates of eq.~\eqref{eq.logistic} (trajectory) must fall along $f(x)$ curve (parabola). Thin red line in the left panel shows, how to construct the trajectory geometrically for initial $x_0$. The middle and right panels illustrate the appearance of the period doubling and tangent bifurcations, respectively. In all panels filled and open circles correspond to stable and unstable fixed points, respectively. For clarity, in the right panel these are plotted only for $k=3.84$.}
\label{fig.log_bifrm}
\end{figure*}

The consecutive iterates of eq.~\eqref{eq.logistic} may be constructed geometrically, using the plot of $f(x)$ vs. $x$ as illustrated in the left panel of Fig.~\ref{fig.log_bifrm} for $k=2.6$ (thin red line). For initial value $x_0$ one plots the vertical line towards the $f$ and from that point the horizontal line towards the diagonal, and then repeats the procedure to get a trajectory. In the discussed case it converges to the fixed point, $a_1=f(a_1)$ (period-1 cycle), which is located at the crossing of $f$ and the diagonal. Its stability depends on the slope of $f$ at the intersection. If the modulus of the slope is less than 1 the iterations converge towards $a_1$ and the period-1 cycle is stable. It is the case for $1<k\le 3$: iterations of any  $x_0$ converge towards $a_1=1-1/k$. For $k>3$ the slope is steeper than 1, the period-1 cycle becomes unstable and stable period-2 cycle ($a_2=f^2(a_2)$) is born through the period doubling (pitchfork) bifurcation. The iterates of eq.~\eqref{eq.logistic} alternate between two values. The appearance of the period doubling bifurcation is illustrated in the middle panel of Fig.~\ref{fig.log_bifrm} with the help of $f^2(x)$ vs. $x$ plot. For $k<3$ $f^2$ intersects the diagonal at a single point corresponding to a stable period-1 cycle. At the bifurcation point ($k_1=3$) $f^2$ is tangent to the diagonal, and for larger $k$, $f^2$ intersects the diagonal at three points. The middle point corresponds to the unstable period-1 cycle, which is a degenerate case of period-2 solution. The two other points have the same slopes and correspond to the stable period-2 cycle. 

\begin{figure}
\centering
\resizebox{\hsize}{!}{\includegraphics{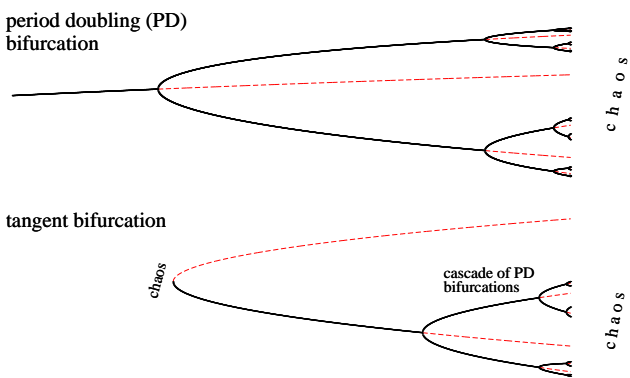}}
\caption{Schematic illustration of the period doubling cascade (top) and tangent bifurcation followed by cascade of period doubling bifurcations (bottom). Black solid and red dashed curves correspond to stable and unstable cycles, respectively.}
\label{fig.log_scheme}
\end{figure}

Following the same scenario, at $k_2=1+\sqrt{6}$ another period doubling bifurcation occurs giving rise to stable period-4 cycle (and period-2 cycle loses its stability). The cascade of period doubling bifurcations will finally lead to chaos. The extent of the domain of stable period-$n$ cycle ($d_n=k_n-k_{n/2}$) decreases as $n$ increases and the ratio $d_{n}/d_{2n}$ approaches the Feigenbaum constant ($\delta\approx 4.669$) which is a universal constant for many other systems too \citep[Feigenbaum's universality, see][]{feigenbaum}. Beyond $k_\infty\approx 3.569946$ (the accumulation point) chaos appears for the first time. The described period doubling route to chaos is schematically illustrated in  the upper part of Fig.~\ref{fig.log_scheme} with stable and unstable cycles marked with black solid and red dashed lines, respectively. 

Beyond the accumulation point the chaotic domain extends, which however, is densely packed with windows of order -- stable period-$n$ cycles. To understand how these are born we focus our attention on the most prominent period-3 window (zoomed in the right part of Fig.~\ref{fig.log_bifurcation}) and the iteration of $f^3$ (rightmost panel of Fig.~\ref{fig.log_bifrm}). Before the period-3 cycle is born the system is chaotic and the three vertices of $f^3$ approach the diagonal. At/beyond the bifurcation ($k=1+2\sqrt{2}$) they touch/intersect the diagonal and give rise to a pair of period-3 cycles, of which one is stable and one is unstable, as is apparent from the analysis of the slopes of $f^3$ at the intersection with the diagonal. The stable period 3-cycle will soon undergo a series o period doubling bifurcations en route to chaos, just as described in the previous paragraphs, and as is clearly visible in the right panel of Fig.~\ref{fig.log_bifurcation}. This scenario  is schematically plotted in the bottom part of Fig.~\ref{fig.log_scheme}.

\begin{figure*}
\centering
\resizebox{\hsize}{!}{\includegraphics{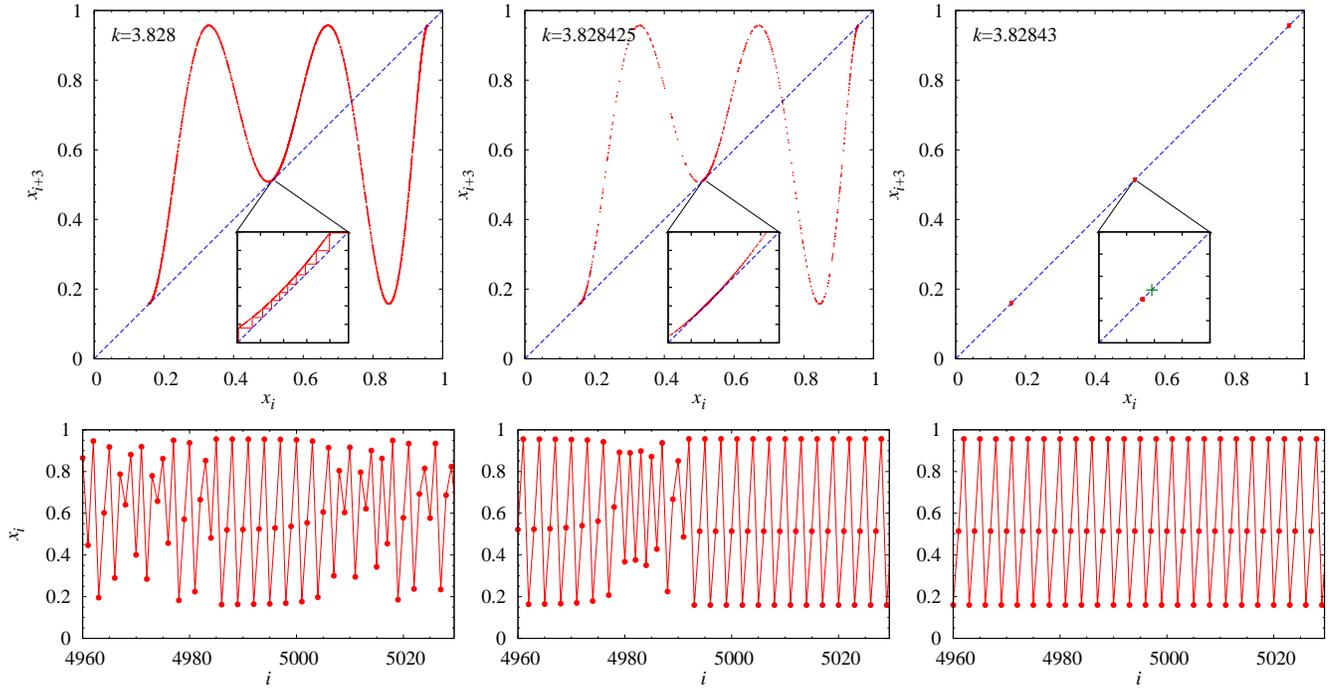}}
\caption{Demonstration of type-I intermittency. The bottom panels show consecutive iterations of the logistic equation for three different values of $k$. The consecutive iterates are connected with lines for guidance. In the top panels the corresponding third return maps are plotted. In the top right panel (inset) the location of unstable fixed point of period-3 is indicated with $+$.}
\label{fig.log_intermittency}
\end{figure*}

\begin{figure*}
\centering
\resizebox{\hsize}{!}{\includegraphics{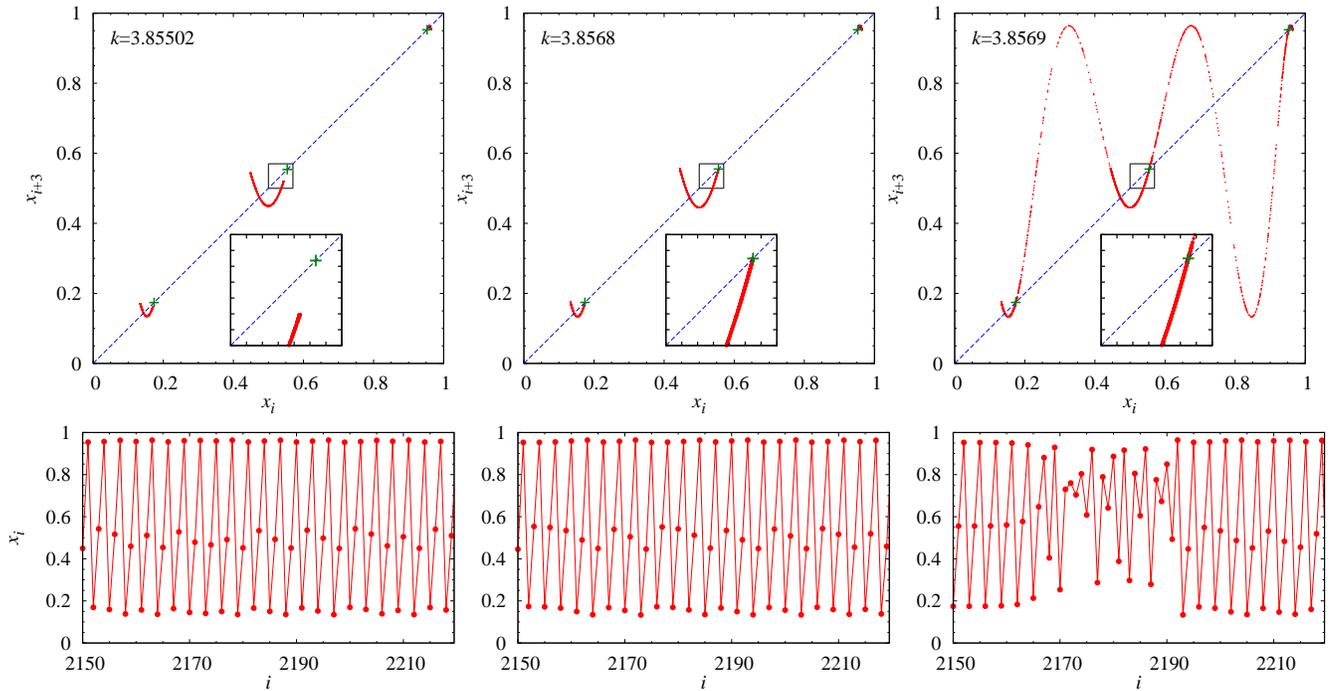}}
\caption{Demonstration of interior crisis. The bottom panels show consecutive iterations of the logistic equation for three different values of $k$. The consecutive iterates are connected with lines for guidance. In the top panels the corresponding third return maps are plotted. '+' signs mark the location of the unstable fixed points of period-3 born through the tangent bifurcation. See also right panel of Fig.~\ref{fig.log_bifurcation}, dashed lines.}
\label{fig.log_crises}
\end{figure*}

Two interesting phenomena occur at the edges of the just discussed period-3 window (and other periodic windows as well). Before the tangent bifurcation occurs we observe the intermittency, illustrated in Fig.~\ref{fig.log_intermittency}. The bottom panels show the consecutive iterations of eq.~\eqref{eq.logistic} for three different values of $k$ and the top panels show the third return maps, i.e. plots of $x_{i+3}$ vs. $x_i$. Obviously the points fall along the $f^3$ curve. Just before the bifurcation (left and middle panel) evolution of the system is characterized by long intervals of almost periodic behaviour interrupted by shorter bursts of chaos. This is the intermittent behaviour first analysed for the Lorenz equations by \cite{MP79}, followed by an in-depth analysis by \cite{PM80}. As control parameter increases the almost periodic intervals become longer (cf. left and middle panels in Fig.~\ref{fig.log_intermittency}) up to a critical value of $k$ at which tangent bifurcation occurs and stable period-3 cycle is born (and unstable period-3 cycle as well; right panel in Fig.~\ref{fig.log_intermittency}).

To discuss the appearance of intermittency we focus attention on the left panel of Fig.~\ref{fig.log_intermittency} (inset). Constructing the trajectory for each third iterate of eq.~\eqref{eq.logistic} geometrically, one must fall into the intermittent channel -- a narrow region between the vertex of $f^3$ and the diagonal -- in which iterations must be trapped for a while (thin zig-zag in the inset). Similar channels are also present at the two other vertices approaching the diagonal and iterates of eq.~\eqref{eq.logistic} fall consecutively into the three channels. Closer to the bifurcation, narrower the channels and longer the iterations are trapped within, with apparently more periodic variation (middle panel of Fig.~\ref{fig.log_intermittency}, inset). As the iterations leave the intermittent channels a chaotic burst is observed.

In a broader context, intermittency is one of the routes to chaos \citep[for a review see][]{eckmann}, characterized by sporadic switching between qualitatively different behaviours, the laminar (periodic) behaviour and chaotic bursts. Intermittency is associated with a bifurcation in which stable periodic cycle becomes unstable. Depending on the bifurcation in which the stability is lost, three types of intermittency were distinguished by \cite{PM80}. The type-I intermittency is related to a tangent bifurcation in which stable and unstable limit cycles collide and both vanish. It is the case for the just described intermittency in the logistic map. Type-II intermittency is related to the subcritical Hopf bifurcation (in the Hopf bifurcation the stationary solution bifurcates into periodic orbit)\footnote{We follow the convention adopted e.g. in \cite{book.seydel} and call the bifurcation subcritical if the stable solution exists on one side of the bifurcation point only. If the stable solution exists on either side, the bifurcation is supercritical.}. Type-III intermittency appears together with subcritical period doubling bifurcation, in which the stable limit cycle collides with the unstable period-doubled cycle. A summary of these scenarios may be found in the original paper by \cite{PM80} and e.g. in \cite{becker}. Hydrodynamic models to be discussed in the coming sections display both type-I and type-III intermittency.

The stable period-3 cycle born in the tangent bifurcation will soon undergo a series of period doubling bifurcations which will create three separate chaotic bands as is well visible in Fig.~\ref{fig.log_bifurcation} and in Fig.~\ref{fig.log_crises} (left and middle panels), which is the same as Fig.~\ref{fig.log_intermittency} except we focus on the right edge of the period-3 window. These three bands will merge in the interior crisis bifurcation, first described by \cite{GOY82}, in which the volume of the attractor changes suddenly. The crisis bifurcation occurs as the three chaotic bands hit the unstable period-3 cycle (born in the tangent bifurcation) which scatters the trajectories into previously unvisited regions -- Fig.~\ref{fig.log_crises} the rightmost panel. This is well visible in the right panel of Fig.~\ref{fig.log_bifurcation} in which unstable cycles of period 3 (and period 6) are plotted with the dashed lines and in Fig.~\ref{fig.log_crises} where location of unstable fixed points of period-3 is indicated with '+' signs.

Directly after the occurrence of crisis bifurcation, crisis induced intermittency is observed (well visible in the bottom right panel of Fig.~\ref{fig.log_crises}). The trajectory is confined in the region of the former, pre-crisis attractor, with sporadic excursions out of it.

%%%%%%%%%%%%%%%%%%%%%%%%%%%%%%%%%%%%%%%%%%%%%%
\section{Hydrodynamic models}\label{sec.hydro}
%%%%%%%%%%%%%%%%%%%%%%%%%%%%%%%%%%%%%%%%%%%%%%

\begin{figure}
\centering
\resizebox{\hsize}{!}{\includegraphics{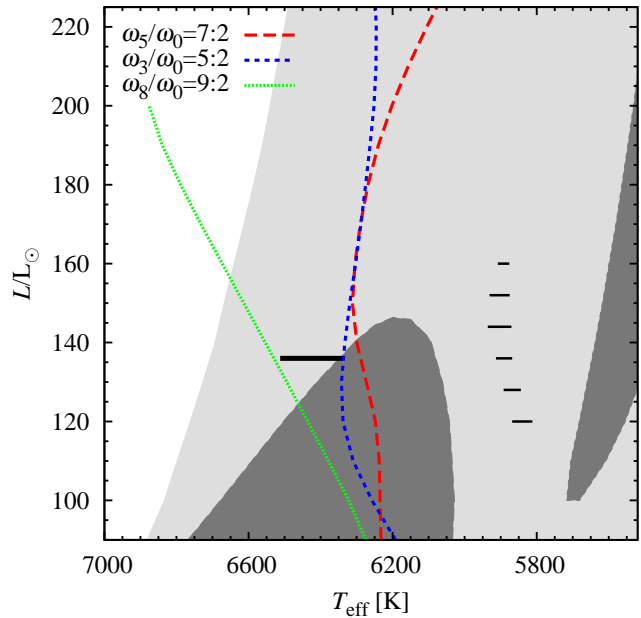}}
\caption{The H-R diagram with the location of models showing chaotic behaviour (thick horizontal line). For reference, models showing periodic and quasi-periodic modulation of pulsation analysed by Smolec \& Moskalik (2012) are marked with thin line segments. The light- and dark-shaded areas mark the fundamental and first overtone instability strips, and dashed lines show the loci of some half-integer resonances running close to the analysed models.}
\label{fig.hr}
\end{figure}

All hydrodynamic models analysed in this paper were computed with the Warsaw non-linear convective pulsation codes \citep{sm08}. Numerical parameters of the models (zoning) are the same as in our previous papers (section 3 of \cite{ssm12} and \cite{sm12}). The physical parameters of the models and parameters of the turbulent convection model \citep{kuhfuss} are the same as in \cite{sm12}. In particular $M=0.55\MS$, $X=0.76$ and $Z=0.0001$. We focus on a single sequence of models with the same luminosity, $L=136\LS$ and varying effective temperature, $T_{\rm eff}$, which is our {\it control} parameter through this paper. The models cover a strip extending over $170$\thinspace K, from $6340$\thinspace K to $6512$\thinspace K (corresponding periods of the fundamental mode are $1.696$\thinspace d and $1.527$\thinspace d). The maximum temperature difference between the consecutive models is only $1$\thinspace K and in the most interesting domains the difference is as small as $0.1$\thinspace K. In Fig.~\ref{fig.hr} we show the location of our models in the H-R diagram (thick horizontal line), together with the location of models that we have studied in \cite{sm12} that show periodic and quasi-periodic modulation of pulsation (thin horizontal lines).

In non-linear computations, the initial static model was perturbed with the velocity profile, and was integrated for at least $10\,000$ pulsation cycles with $1\,200$ time steps per pulsation cycle. By default, all the models were initialized in the same manner, with velocity eigenfunction of the fundamental mode, scaled to match the 4.5\thinspace km\thinspace s$^{-1}$ surface velocity. To check for the possible dependence of results on the initialization, in particular to check whether e.g. two attractors are possible for the same model (hysteresis), several models were initialized in a different manner (with larger surface velocity or with a mixture of the fundamental mode and first overtone eigenfunctions). In all considered cases, the computations converged to the same attractor, only the length of the initial transient phase was different. To check the long-term stability of the computed attractors (either periodic or chaotic), we have computed up to $50\,000$ pulsation cycles for a few models. In all cases the attractor emerging from the calculation of first $10\,000$ pulsation cycles remained stable.

\begin{figure*}
\centering
\resizebox{\hsize}{!}{\includegraphics{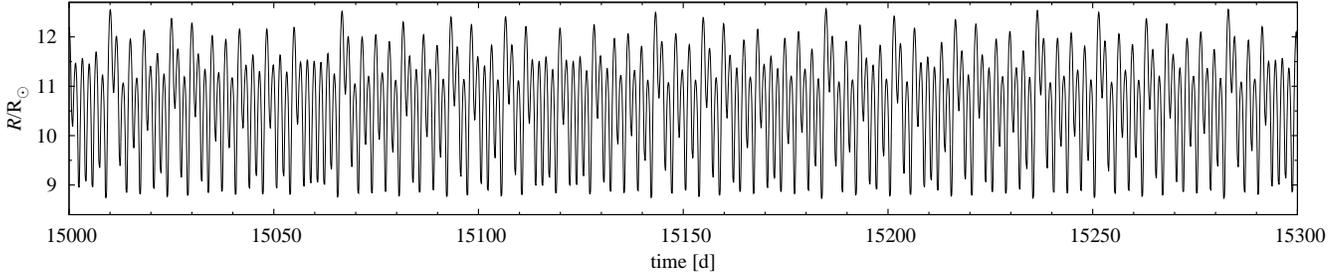}}
\caption{Section of time series for $6410.0$\thinspace K model showing chaotic variability.}
\label{fig.ts_6410}
\end{figure*}

We note that in these models the eddy-viscous dissipation is strongly decreased, $\alpha_m=0.05$ \citep[for equations see][]{sm08}. It results in significant pulsation amplitudes and, as our treatment of radiation is very simple (diffusion approximation) and model mesh is fixed, in erratic light curves with spurious spikes (see section 4 in \cite{sm12} for detailed discussion of this point). Therefore in this paper we analyse the radius variation only, which is smooth. In Fig.~\ref{fig.ts_6410} we plot a section of typical time series for model showing chaotic variability ($6410.0$\thinspace K).

%%%%%%%%%%%%%%%%%%%%%%%%%%%%%%%%%%%%%%%%%%%%%%
\section{Analysis of the BL~Her hydrodynamic models}\label{sec.analysis}
%%%%%%%%%%%%%%%%%%%%%%%%%%%%%%%%%%%%%%%%%%%%%%

In this Section we analyse the radius variation of our hydrodynamic models.

\subsection{Bifurcation diagram}\label{ssec.bifurcation}
%%%%%%%%%%%%%%%%%%%%%%%%%%%%%%%%%%%%%%%%%%%%%%%%%%%%%%%%%%%%%%%%%%%%%%%%%%%%

\begin{figure*}
\centering
\resizebox{\hsize}{!}{\includegraphics{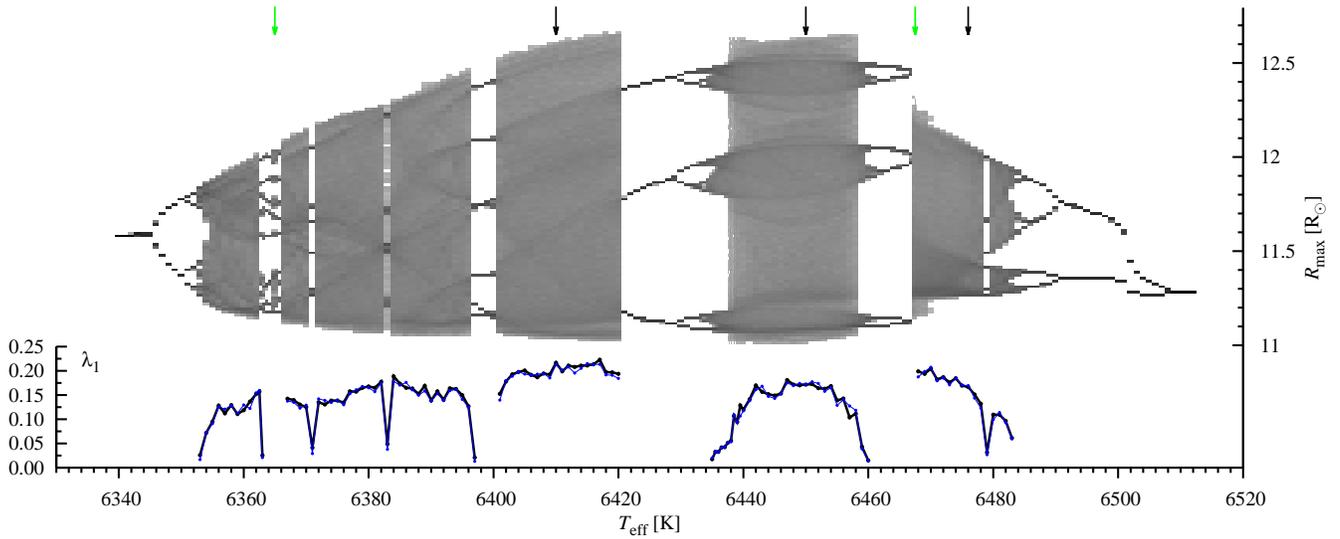}}
\caption{Bifurcation diagram for hydrodynamic BL~Her models constructed with values of maximum radii over $9\,000$ pulsation cycles. Arrows in the top section point the location of particular chaotic (black arrows) and periodic (green arrows) models discussed in Sections~\ref{ssec.cs1} and \ref{ssec.cs2}, respectively. In the bottom we plot the largest Lyapunov exponents for the chaotic models -- see Section~\ref{sec.lyap} for details.}
\label{fig.histo}
\end{figure*}

Deterministic chaos is present in our models beyond doubt. The conclusion is unavoidable once the bifurcation diagram is plotted -- Fig.~\ref{fig.histo}. The diagram is constructed in a similar way as in the case of logistic map (Fig.~\ref{fig.log_bifurcation}). For each value of our control parameter -- the effective temperature -- we computed the probability that the maximum radii, $R_{\rm max}$, falls into one of the $120$ bins into which the range of radius variation in our models ($\sim 11.0-12.7\RS$) was divided. The stack of grey-scaled histograms is displayed in Fig.~\ref{fig.histo} and shows a striking similarity to classical chaotic systems. One can pick other parameter than maximum radius to construct the bifurcation diagram, but results are qualitatively the same. As compared to Fig.~\ref{fig.log_bifurcation} the bifurcation diagram looks rough which results from smaller resolution in the control parameter and decreased number of bins along vertical axis, necessary to have a reasonable statistics in each bin.

On both sides of the computation domain single-periodic (cycle-1) pulsation in a fundamental mode is present. The chaotic bands are reached through a series of period doubling bifurcations both from the cool and the hot side. Qualitatively the same scenario is observed for the Gauss (or {\it mouse map}), see e.g. \cite{hilborn}. The chaotic bands are separated with periodic windows of order. The largest period-3 window is nearly a one-to-one copy of the just discussed counterpart seen in case of the logistic map. Within chaotic bands pronounced structures are clearly visible as well -- the probability of hitting particular bins by maximum radii is not equal. Some values of maximum radii are clearly preferred as indicated with darker bands, migrating across the bifurcation diagram as control parameter changes.

\subsection{Period doubling cascade en route to chaos}\label{ssec.PD}
%%%%%%%%%%%%%%%%%%%%%%%%%%%%%%%%%%%%%%%%%%%%%%%%%%%%%%%%%%%%%%%%%%%%%

Our hydrodynamic models display two pronounced period doubling cascades that lead from a single-periodic fundamental mode pulsation (period-1 cycle) to chaos. The first cascade extends on the cool side of the computation domain. We clearly observe the appearance of period-2 cycle (first in the $6344.0$\thinspace K model and present up to $6350.0$\thinspace K model) and period-4 cycle (first in the $6351.0$\thinspace K model and present also in the $6352.0$\thinspace K model). Further periodic cycles are not resolved in our model grid with $1$\thinspace K step in effective temperature. The maximum radii of the $6353.0$\thinspace K model are bounded within eight well separated chaotic bands, which, as effective temperature is increased, merge smoothly into one chaotic band (at $6355.0$\thinspace K).

The other period doubling cascade, on the hot side of the computation domain, extends over much larger temperature range, but otherwise it is a mirror image of the just discussed cascade. The consecutive period doubling bifurcations occur with decreasing control parameter. For such situation, the terms period-halving or inverse period doubling cascade are in use. Here we describe the route from order (period-1 cycle) to chaos as effective temperature is decreased. Period-2, period-4, period-8 and period-16 cycles are all clearly detected. They appear for the first time in models with temperatures $6508.0$\thinspace K, $6490.0$\thinspace K, $6485.0$\thinspace K and $6484.0$\thinspace K, respectively. Period-8 and period-16 variation is detected in only one model each. In $6483.0$\thinspace K model maximum radii are bounded within four separate chaotic bands. 

It is interesting to check whether the lengths of the domains of the consecutive period-$n$ cycles follow the Feigenbaum scaling. Our model grid is too coarse to exactly pinpoint the bifurcation points and hence only a rough estimates are possible. Assuming that the bifurcation occurs halfway between the neighbouring period-$n$ and period-$2n$ models we get (for the cascade on the hot side): $d_2=18$\thinspace K, $d_4=5$\thinspace K and $d_8=1$\thinspace K for the extents of the period-2, period-4 and period-8 domains, respectively (see Section~\ref{sec.logistic}). The ratios (assuming $0.5$\thinspace K error in the estimate of the bifurcation point) are $d_2/d_4=(3.6\pm 0.4)$\thinspace K and $d_4/d_8=(5\pm 2.5)$\thinspace K not significantly different from the Feigenbaum constant ($\delta\approx 4.669$) toward which the ratio $d_{n}/d_{2n}$ converges as $n\rightarrow\infty$ for the logistic map and other iterated maps, too \citep{feigenbaum,cel80,cek81}. Feigenbaum scaling is also observed with periodic solutions of the ordinary differential equations \citep{book.seydel}. For the cascade on the cool side we may only estimate $d_2/d_4$ which is $(3.5\pm 0.9)$\thinspace K.

\subsection{Case studies 1: chaotic models}\label{ssec.cs1}
%%%%%%%%%%%%%%%%%%%%%%%%%%%%%%%%%%%%%%%%%%%%%%%%%%%%%%%%%%%

Before we describe the phenomena that shape the bifurcation diagram, we present a more detailed analysis for three chaotic models ($6410.0$\thinspace K, $6450.0$\thinspace K and $6476.0$\thinspace). Their location in the bifurcation diagram (Fig.~\ref{fig.histo}) is shown with black arrows. Nearly all of the computed chaotic models were analysed in the same manner as presented below.

\begin{figure*}
\centering
\resizebox{\hsize}{!}{\includegraphics{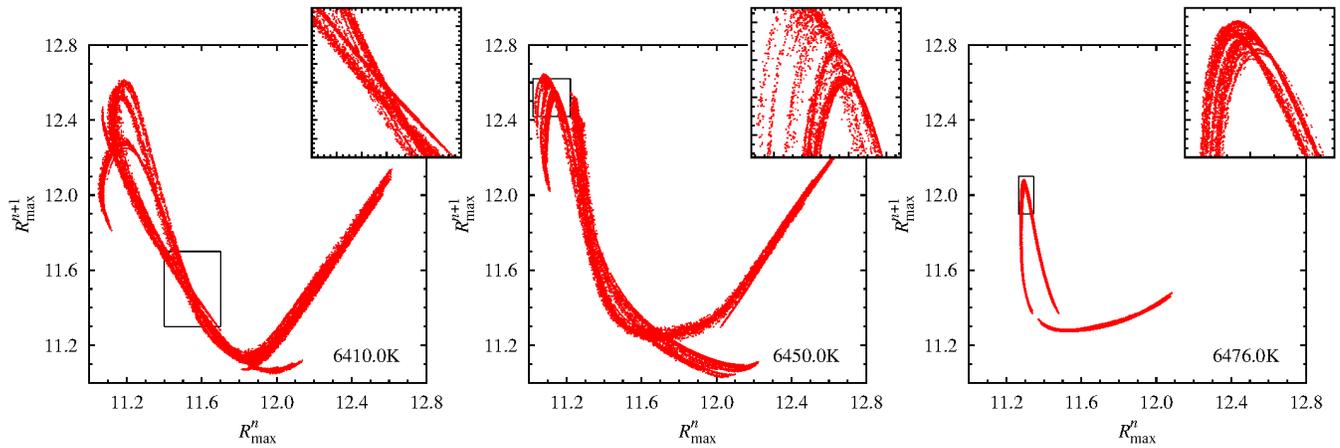}}
\caption{First return maps for three chaotic models (marked with arrows in Fig.~\ref{fig.histo}). Data for the last $46\,000$ cycles (out of $50\,000$ computed) are plotted. {\it[This is low resolution version of the Figure]}}
\label{fig.rm_chaos}
\end{figure*}

There is not much to learn from the time-series alone. As an example in Fig.~\ref{fig.ts_6410} we show radius variation over $\approx\!180$ pulsation cycles for $6410.0$\thinspace K model, and indeed, no obvious regularity can be noticed. Much more useful are return maps for maximum radii, ie. plots of $R_{\rm max}^{n+1}$ vs. $R_{\rm max}^{n}$. These are Poincar\'e maps with surface of section defined by $dR/dt=0$ at maximum expansion phase. The maps are shown in Fig.~\ref{fig.rm_chaos} for the discussed models. The points do not populate the plot in a random manner but fall along a characteristic, albeit rather complex shape, which evolves as the effective temperature changes. This evolution is illustrated with animation that may be found in the online version of this article as additional supporting information (see also other maps for chaotic models of different effective temperatures -- grey dots in Figs.~\ref{fig.rm_9} and \ref{fig.rmpw}). The complex shape, as compared to analogous maps for some classical systems (e.g. tent map for the maximum $z$ values in the Lorenz system, see \cite{lorenz} or in \cite{book.PJS}) is not surprising. Our system is much more complex and return map is only a 2D projection of complex dynamics occurring on a higher dimension manifold. The insets in Fig.~\ref{fig.rm_chaos} provide insight into the fine-structure of the chaotic attractor. Although the numerical noise does not allow to show the cascade of such zoom-ins into smaller and smaller regions we conclude that the attractor's structure is most likely fractal and the attractor is strange.

\begin{figure}
\centering
\resizebox{\hsize}{!}{\includegraphics{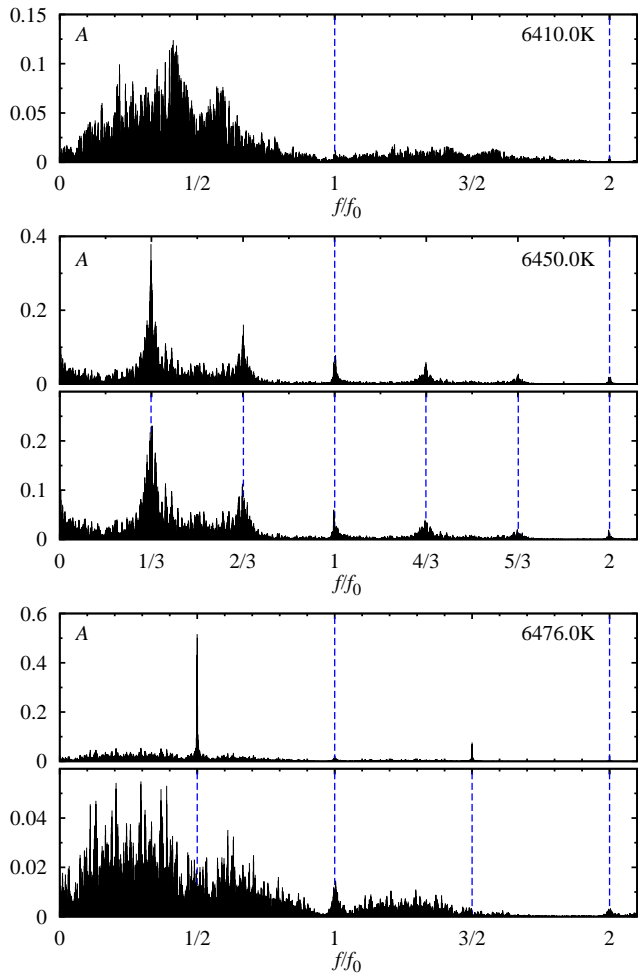}}
\caption{Fourier spectra for three models of different effective temperature showing chaotic behaviour. Dashed lines show the location of pre-whitened frequencies (fundamental mode and harmonics in each case, and additional frequencies for the $6450.0$\thinspace K and $6476.0$\thinspace K models.}
\label{fig.fs_chaos}
\end{figure}

Chaos clearly manifests in the Fourier spectra, which are plotted in Fig.~\ref{fig.fs_chaos}. In each case a time series for $\approx 650$ pulsation cycles was analysed with \textsf{Period04} software \citep{period04}. The spectra were pre-whitened with the frequency of the fundamental mode and its harmonics -- location of these frequencies is shown with dashed lines. For two models also additional highest-signal frequencies in the resulting spectra were pre-whitened and the result is also shown in Fig.~\ref{fig.fs_chaos}.  

In each case wide bands of signal are present in the spectra, which are characteristic signature of chaos. One large band is present for the $6410.0$\thinspace K model (Fig.~\ref{fig.fs_chaos}, top panel), without any obvious structure. To the contrary, in the case of the $6450.0$\thinspace K model (Fig.~\ref{fig.fs_chaos}, middle panel), wide bands are concentrated around $1/3f_0$ and its harmonics. The cause of this difference becomes clear once the bifurcation diagram is analysed (Fig.~\ref{fig.histo}). In the case of $6410.0$\thinspace K model there are no obvious preferred values, or ranges of values, for the maximum radii. For the $6450.0$\thinspace K model, the maximum radii fall preferentially into three ranges which manifest in Fig.~\ref{fig.histo} as three dark-grey bands within one large chaotic domain extending between $6438.0$\thinspace K and $6458.0$\thinspace K. The signal at $1/3f_0$ and its harmonics is obviously not coherent which successive pre-whitening (see Fig.~\ref{fig.fs_chaos}) shows.

The situation is slightly different for the $6476.0$\thinspace K model (Fig.~\ref{fig.fs_chaos}, bottom panel). After pre-whitening with the fundamental mode and its harmonics strong signal is present at subharmonic frequencies. It results from the two-band structure of the attractor clearly visible in the return map (Fig.~\ref{fig.rm_chaos}, right panel). The maximum radii alternate between the two bands and vary chaotically within each of them. Therefore, signal at $1/2f_0$ and its harmonics is strong and highly coherent but, after pre-whitening with these frequencies, only very wide bands of signal without any obvious structure are present in the frequency spectra.

\subsection{Periodic windows of order}\label{ssec.PW}
%%%%%%%%%%%%%%%%%%%%%%%%%%%%%%%%%%%%%%%%%%%%%%%%%%%%%

\begin{figure}
\centering
\resizebox{\hsize}{!}{\includegraphics{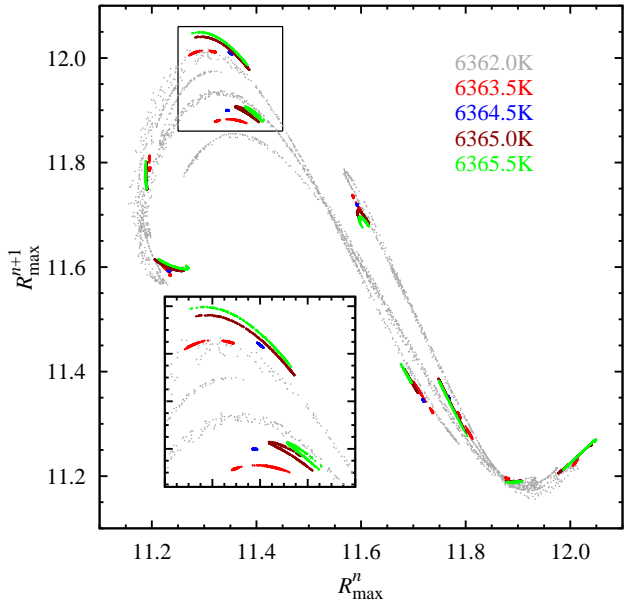}}
\caption{First return map for four models located within period-9 window. For a reference, the return map for the closest and cooler chaotic model is plotted with grey dots. Data for the last $3\,000$ pulsation cycles out of the total $10\,000$ computed cycles are plotted.}
\label{fig.rm_9}
\end{figure}

\begin{figure}
\centering
\resizebox{\hsize}{!}{\includegraphics{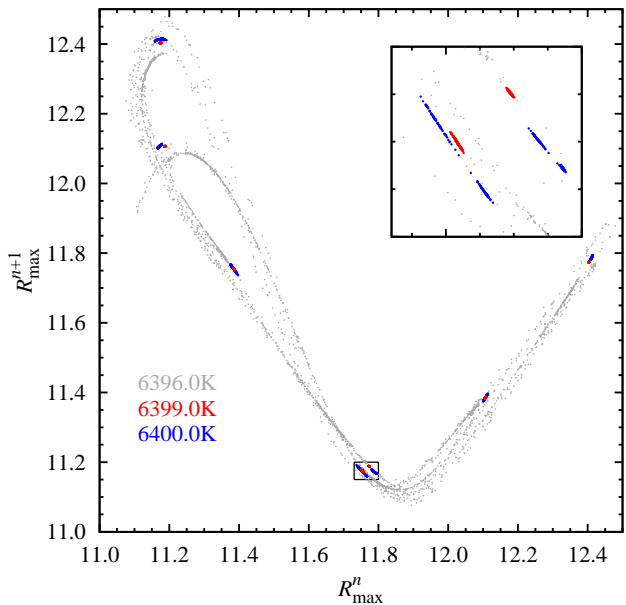}}
\caption{First return map for two models located within period-7 window. For a reference, the return map for the closest and cooler chaotic model is plotted with grey dots. Data for the last $3\,000$ pulsation cycles out of the total $10\,000$ computed cycles are plotted.}
\label{fig.rm_7}
\end{figure}

\begin{figure*}
\centering
\resizebox{\hsize}{!}{\includegraphics{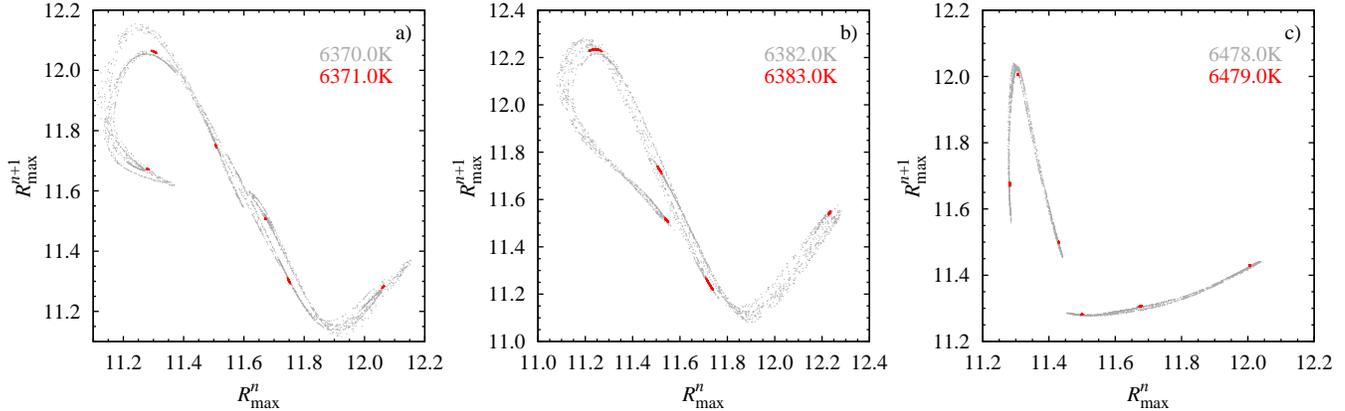}}
\caption{First return maps for models within different periodic windows: a) period-6 window, b) period-5 window and c) period-6 window discussed in Section~\ref{ssec.PW}. In each case the the return map for the closest cooler chaotic model is plotted with light-grey dots for reference. In each case data for the last $3\,000$ pulsation cycles out of the total $10\,000$ computed cycles are plotted.}
\label{fig.rmpw}
\end{figure*}

Within the chaotic band (Fig.~\ref{fig.histo}) seven windows, in which much more ordered behaviour is observed, including strictly periodic variation, may be identified. Below they are briefly described in the order they appear as effective temperature is increased. The return maps for some of the models are plotted in Figs.~\ref{fig.rm_9}, \ref{fig.rm_7} and \ref{fig.rmpw}. In each case the neighbouring chaotic model (of lower effective temperature) is plotted with grey points for reference. The most interesting period-3 window is discussed in more detail in Section~\ref{ssec.intercrises} and two interesting models from period-6 and period-9 windows are discussed in Section~\ref{ssec.cs2}

\begin{figure}
\centering
\resizebox{\hsize}{!}{\includegraphics{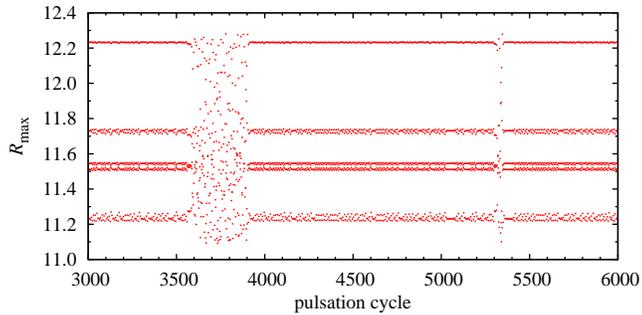}}
\caption{Values of maximum radii plotted over $3\,000$ pulsation cycles for $6383.0$\thinspace K model showing a complex variation with chaotic bursts (at pulsation cycles between $3\,600-3\,900$ and at $5\,800$).}
\label{fig.6383LOG}
\end{figure}

\begin{itemize}
\item[] {\bf Period-9} window is present for models in a range $6363.0-6366.0$\thinspace K (Fig.~\ref{fig.rm_9}). In fact this window is not strictly periodic but display a complicated internal structure. Within this window models were computed with a smaller 0.5\thinspace K-step in effective temperature. For all models nine bands are clearly visible, which are either very wide ($6365.0$\thinspace K, $6365.5$\thinspace K) or very narrow ($6364.5$\thinspace K) indicating a possible strict period-9 cycle. In the case of $6363.5$\thinspace K model each of the nine bands is split and thus 18 bands are apparent. A detailed analysis of the $6365.0$\thinspace K model (that shows a rare case of type-III intermittency) is presented in Section~\ref{ssec.cs2}.

\item[] {\bf Period-6} window is present for one model of $6371.0$\thinspace K (Fig.~\ref{fig.rmpw}a). The neighbouring $\pm 1$\thinspace K models display one chaotic band. Certainly our model grid lacks resolution to provide more insight into the dynamic scenarios within such a narrow window. In return map six very small bands rather than points are present, and thus model is not strictly periodic. The periodic model might be located just a tiny fraction of Kelvin away. This remark also applies to other very narrow windows discussed below.

\item[] {\bf Period-5} window is present for one model of $6383.0$\thinspace K (Fig.~\ref{fig.rmpw}b). The neighbouring $\pm 1$\thinspace K models display one chaotic band. Detailed analysis shows that in fact for this model the maximum radii form five narrow chaotic bands which may join during the {\it chaotic bursts} for several pulsation cycles. Two such bursts happened within $10\,000$ cycle integration of the model and are illustrated in Fig.~\ref{fig.6383LOG}. In this model we also detect a signature of type-III intermittency: switching between period-5 cycle and period-10 cycle. Since the effect is barely visible for this model, we postpone its discussion to Section~\ref{ssec.cs2}, in which we present a more clear example of type-III intermittency in one model in period-9 window.

\item[] {\bf Period-7} window is present for models in a range $6397.0-6400.0$\thinspace K (Fig.~\ref{fig.rm_7}). In-between $6399.0$\thinspace K and $6400.0$\thinspace K period doubling bifurcation occurred. As inset in Fig.~\ref{fig.rm_7} shows, each of the seven, very narrow bands present in the $6399.0$\thinspace K model is split into two wider bands in the slightly hotter, $6400.0$\thinspace K model.

\item[] {\bf Period-3} window with subsequent period doubling cascade extends between $6421.0$\thinspace K (period-3 cycle) and $6438.0$\thinspace K (three chaotic bands). This largest window is discussed in detail in the next Section.

\item[] {\bf Period-6} window with subsequent (inverse) period doubling cascade. The window extends between $6459.0$\thinspace K, at which three chaotic bands are present, till $6468.0$\thinspace K at which period-6 cycle, reached through the inverse period doubling cascade, ceases to exist. This window is in fact a mirror image of the just discussed period-3 window, except the (inverse) period doubling cascade seems to be truncated at the cease of period-6 cycle, which is followed by chaos instead of stable period-3 cycle. The scenario around $6468.0$\thinspace K is uncommon and will be discussed in more detail in Section~\ref{ssec.cs2} and in Section~\ref{sec.concl}. 

\item[] The latter two windows represent a clear example of {\bf period-3 bubble} \citep[or {\it remerging Feigenbaum tree},][]{rft}. The outlook at the bifurcation diagram (Fig.~\ref{fig.histo}) reveals that the two just discussed windows are in fact tightly connected. The scenarios at the cool and hot side of the chaotic band separating the windows and extending between $6438.0-6459.0$\thinspace K are not only mutual mirror images. In fact the three chaotic bands formed at the two sides of the chaotic domain do not disappear as they merge into one chaotic band (in the interior crisis bifurcation, see next Section), but sustain their identity and smoothly merge within the chaotic domain as the dark-grey bands in Fig.~\ref{fig.histo} indicate.

\item[] {\bf Period-6} window is present for one model of $6479.0$\thinspace K (Fig.~\ref{fig.rmpw}c). The neighbouring $\pm 1$\thinspace K models display two chaotic bands. The ones at the cool side merge into one band at $6475.0$\thinspace K.
\end{itemize}

\subsection{Intermittency and crisis at period-3 window}\label{ssec.intercrises}
%%%%%%%%%%%%%%%%%%%%%%%%%%%%%%%%%%%%%%%%%%%%%%%%%%%%%%%%%%%%%%%%%%%%%%%%%%%%%%%%

In this Section we focus attention on the largest period-3 window extending between $6421.0$\thinspace K and $6438.0$\thinspace K, and its direct vicinity. For the most interesting temperature ranges, at the edges of the window, the models were computed with a very small, $0.1$\thinspace K-step in effective temperature. The corresponding part of the bifurcation diagram displays a striking similarity to the bifurcation diagram of the logistic map, Fig.~\ref{fig.log_bifurcation}, and bifurcation diagrams of many other dynamical systems \citep[e.g. R\"ossler system, see in][]{book.PJS}. This similarity, and the analysis presented in this Section, lead to conclusion that in these systems the same dynamical phenomena lead to the appearance of the period-3 window and its subsequent evolution to chaos.

\begin{figure}
\centering
\resizebox{\hsize}{!}{\includegraphics{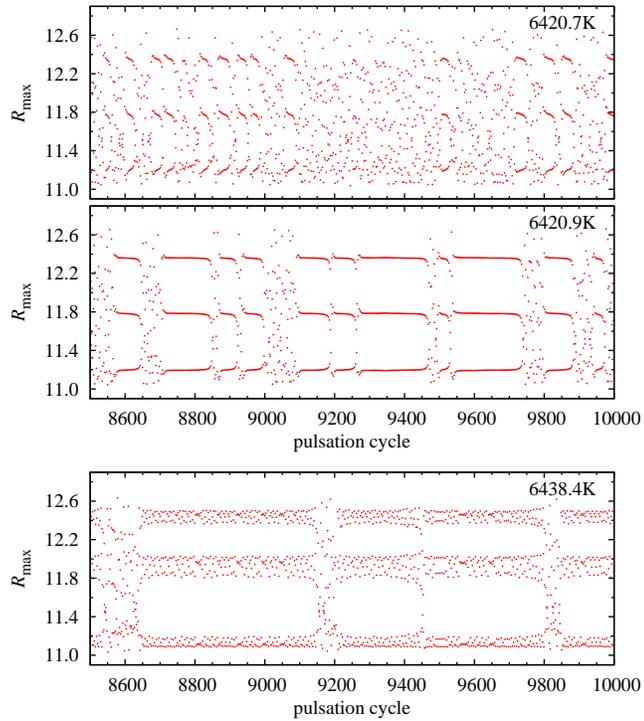}}
\caption{Maximum radii plotted over $1\,500$ consecutive pulsation cycles for three models showing the type-I intermittency (top two panels) and crisis induced intermittency (bottom panel).}
\label{fig.intermittencyLOG}
\end{figure}

\begin{figure*}
\centering
\resizebox{\hsize}{!}{\includegraphics{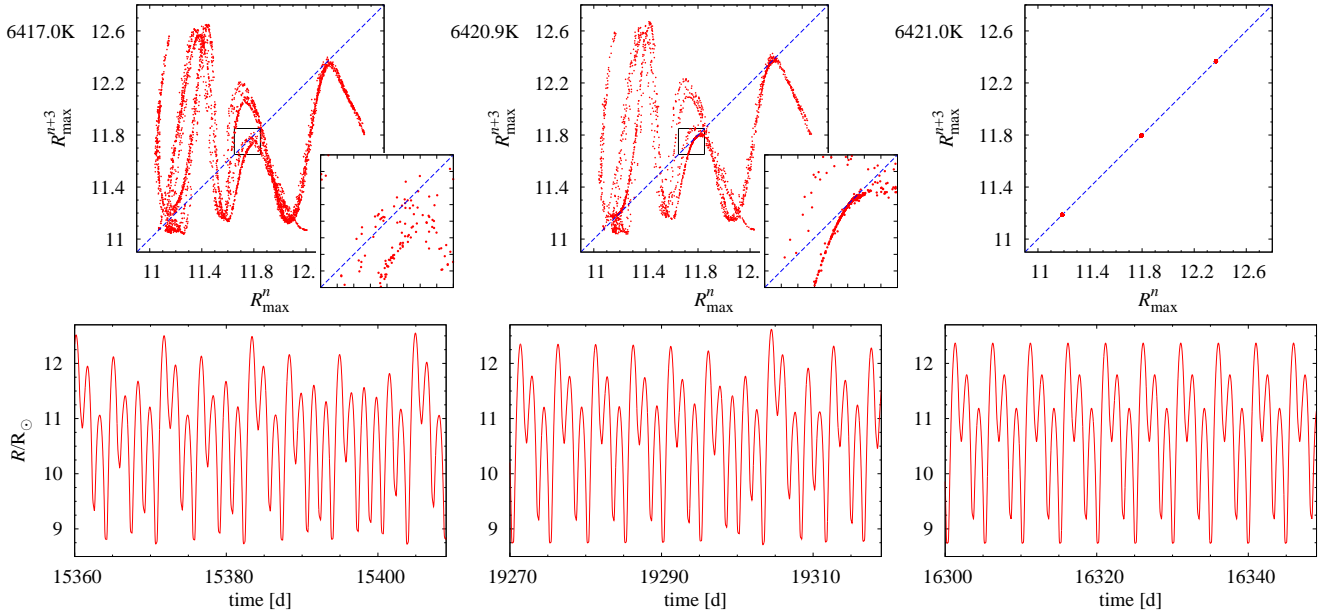}}
\caption{Illustration of type-I intermittency at the onset of period-3 window. The top panels show the third return maps and the bottom panels show the sections of radius variation for three models of different effective temperature. Note that the last two models are separated by $0.1$\thinspace K, only. Compare with Fig.~\ref{fig.log_intermittency} for the logistic map.}
\label{fig.intermittency}
\end{figure*}

In Fig.~\ref{fig.intermittencyLOG} (top two panels) we show the values of maximum radii during the $1\,500$ pulsation cycles in two models directly preceding the appearance of the period-3 window. Intermittency is clearly visible. For the $6420.7$\thinspace K model the intervals with apparently periodic, period-3 variation are short, last by up to 50 pulsation cycles, and are interrupted by much longer intervals of chaos. As the period-3 window is approached the almost periodic behaviour dominates, as is the case for the slightly hotter model of $6420.9$\thinspace K. The long intervals of period-3 behaviour are only sporadically interrupted with much shorter bursts of chaos. This behaviour is characteristic for type-I intermittency. The appearance of intermittency and the birth of the period-3 window are further illustrated in Fig.~\ref{fig.intermittency}. It shows the third return maps (top) and small sections of radius variation for three different models at the onset of period-3 window. The Figure is a counterpart of Fig.~\ref{fig.log_intermittency} for the logistic map. The intermittent behaviour is well visible in the models directly preceding the period-3 window. The return maps show the formation of the intermittent channels (one is zoomed in the insets for $6417.0$\thinspace K and $6420.9$\thinspace K models), which get narrower as effective temperature is increased and bifurcation point is approached. As system evolves through the intermittent channels the apparently periodic variation is observed. For the hotter model, the density of points at the very narrow intermittent channel is high indicating long intervals of periodic variation. This is also illustrated with the sections of radius variation (bottom panels of Fig.~\ref{fig.intermittency}) which were chosen to show the almost periodic variation interrupted with short chaotic burst (at $\approx 19\,296$\thinspace d for $6420.9$\thinspace K model). In between $6420.9$\thinspace K and $6421.0$\thinspace K the intermittent channel touches the diagonal and period-3 cycle is born in the tangent  bifurcation (rightmost panel in Fig.~\ref{fig.intermittency}).

The ensuing scenario follows the one depicted schematically in the lower part of Fig.~\ref{fig.log_scheme} and discussed in more detail for the logistic map (Section~\ref{sec.logistic}). As effective temperature is increased, the period-3 cycle undergoes a series of period doubling bifurcations en route to chaos. The first bifurcation takes place at $\approx 6429.0$\thinspace K and leads to period-6 cycle. The model at $6434.0$\thinspace K is a period-12 cycle. Further period doublings are not resolved in our computations and the following models display a 6-band chaos and finally 3-band chaos.

\begin{figure*}
\centering
\resizebox{\hsize}{!}{\includegraphics{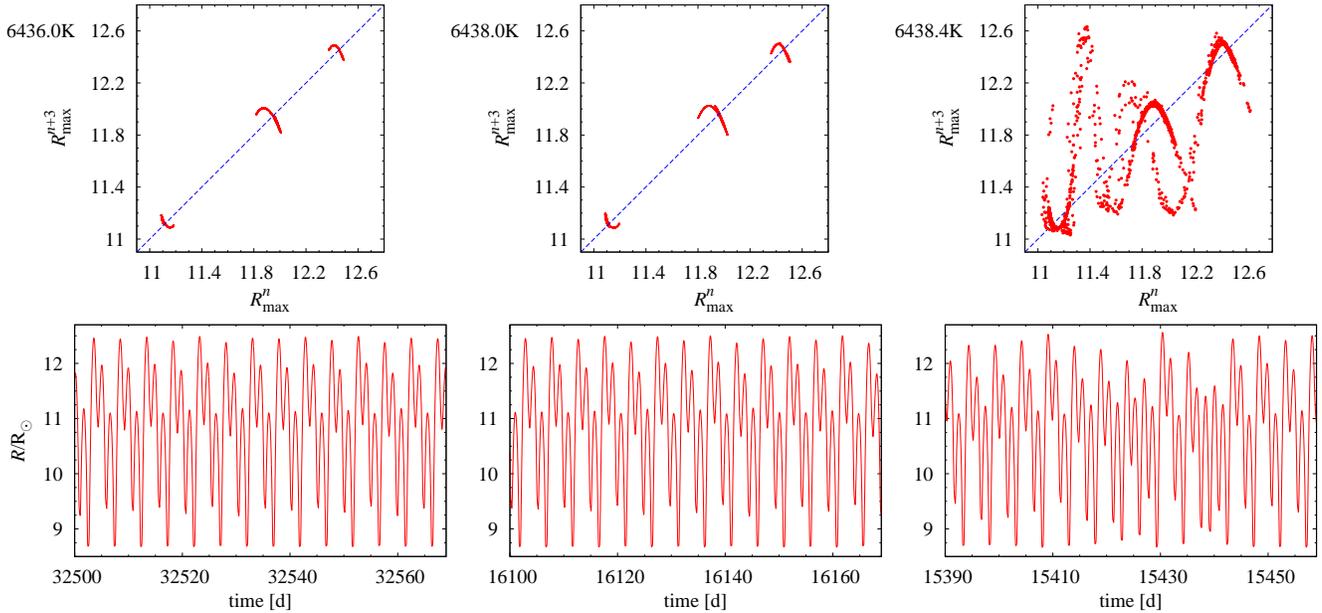}}
\caption{Illustration of interior crisis at the cease of period-3 window. The top panels show the third return maps and the bottom panels show the sections of radius variation for three models of different effective temperature. Compare with Fig.~\ref{fig.log_crises} for the logistic map.}
\label{fig.crises}
\end{figure*}

The three chaotic bands merge in an interior crisis bifurcation which is illustrated with the help of Fig.~\ref{fig.crises}, which should be analysed together with its counterpart for the logistic map -- Fig.~\ref{fig.log_crises}. For the first two models 3-banded chaos is present -- the values of maximum radii are bounded in three, well separated ranges, which is clearly visible both in the return maps and in the sections of radius variation. At the tangent bifurcation leading to the appearance of period-3 window the unstable period-3 cycle was also created. Its exact location in the return map cannot be computed as was the case for logistic map, but the corresponding three points must be located at the diagonal. As effective temperature is increased (cf. left and middle panels of Fig.~\ref{fig.crises}) the three chaotic bands expand, approaching the diagonal and the anticipated unstable period-3 cycle, location of which may now be easily guessed. At the interior crisis bifurcation the three-band chaotic attractor collides with the unstable period-3 cycle and expands into one-band chaotic attractor. 

Directly after the crisis bifurcation, the maximum radii still fall preferentially into the ranges defined by the three chaotic bands. It is well visible in the return map in the rightmost panel of Fig.~\ref{fig.crises} (higher density of points along the three extended arches) and in the corresponding section of radius variation. In addition in Fig.~\ref{fig.intermittencyLOG} (bottom panel) we plot the values of maximum radii during the $1\,500$ consecutive pulsation cycles for the same model. For many pulsation cycles the system evolves in a phase-space defined by the former three-band chaotic attractor and only sporadically gets scattered over a larger space. This behaviour is called {\it crisis induced intermittency} and ceases as effective temperature is increased, albeit, still the probability of maximum radii falling into one of the three bands is larger thorough the full chaotic domain separating the period-3 and period-6 windows ($6438.0-6458.0$\thinspace K; see Fig.~\ref{fig.histo}).

%%%%%%%%%%%%%%%%%%%%%%%%%%%%%%%%%%%%%%%%%%%%%%%%%%%%%%%%%%%%
\subsection{Case studies 2: periodic models}\label{ssec.cs2}
%%%%%%%%%%%%%%%%%%%%%%%%%%%%%%%%%%%%%%%%%%%%%%%%%%%%%%%%%%%%

\begin{figure}
\centering
\resizebox{\hsize}{!}{\includegraphics{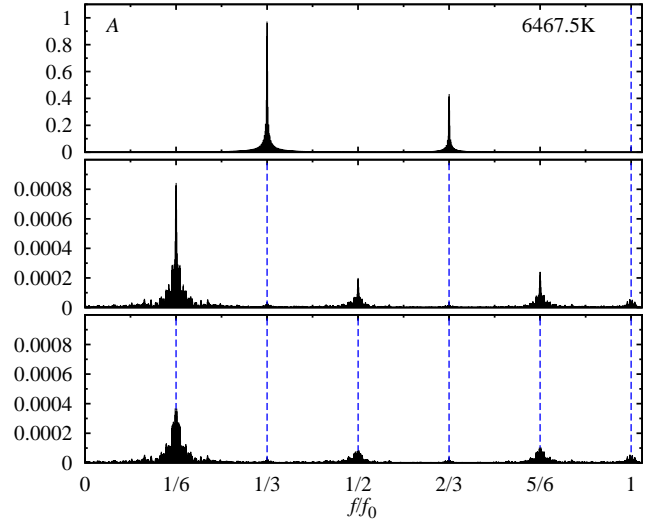}}
\caption{Fourier spectra for $6467.5$\thinspace K model. Dashed lines show the location of pre-whitened frequencies (fundamental mode in the top panel).}
\label{fig.fs_6467.5}
\end{figure}

\begin{figure}
\centering
\resizebox{\hsize}{!}{\includegraphics{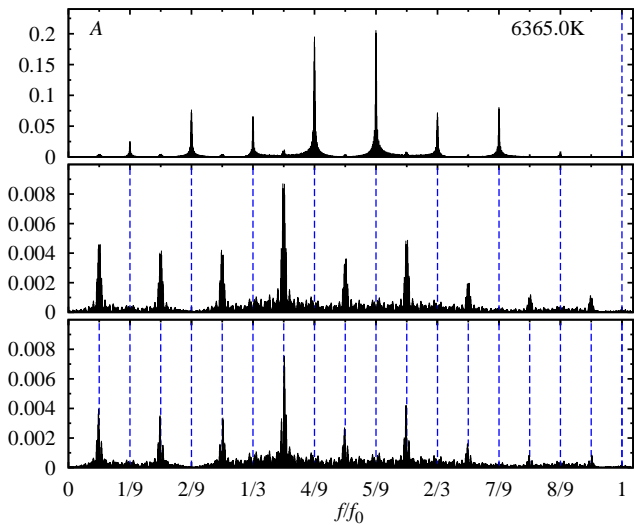}}
\caption{Fourier spectra for $6365.0$\thinspace K model. Dashed lines show the location of pre-whitened frequencies (fundamental mode in the top panel).}
\label{fig.fs_6365.0}
\end{figure}

In this Section we discuss in more detail two models displaying periodic variation: $6467.5$\thinspace K and $6365.0$\thinspace K models. In the bifurcation diagram (Fig.~\ref{fig.histo}) their location is indicated with green arrows. The first model is located within period-6 window while the second is located within period-9 window. Figs.~\ref{fig.fs_6467.5} and \ref{fig.fs_6365.0} show the frequency spectra for the two models. For clarity, these are limited to $[0,f_0]$ range.

\begin{figure}
\centering
\resizebox{\hsize}{!}{\includegraphics{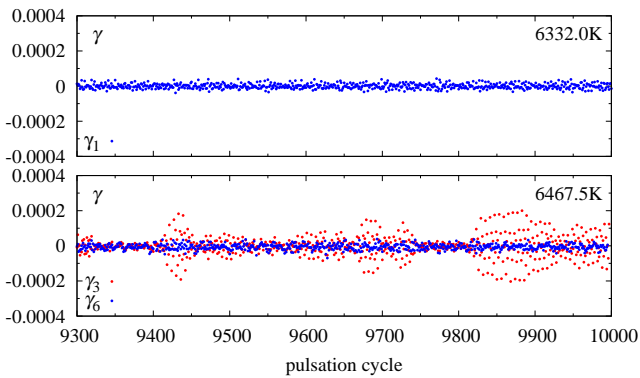}}
\caption{Bottom panel: kinetic energy growth rates, $\gamma_k$, summed over three (red points) or six (blue points) pulsation cycles, for $6467.5$\thinspace K model. In the top panel kinetic energy growth rate for singly-periodic $6332.0$\thinspace K model is plotted for comparison.}
\label{fig.gamma}
\end{figure}

As arrow in Fig.~\ref{fig.histo} indicates  $6467.5$\thinspace K model is located at the border of chaotic band and period-6 window. Whether it is in fact a period-3 cycle variation, just a moment before the period doubling bifurcation, or period-6 cycle, just after the bifurcation, is not clear from the bifurcation diagram. Analysis of return map, which consists of three slightly elongated clumps extending over less than $0.01\RS$ does not provide the clue, either. Before we discuss the Fourier spectrum for the considered model, we briefly describe other method that may be used to resolve the issue, which we find particularly useful for similar models at the direct vicinity of bifurcation points. We analyse growth of the maximum kinetic energy of the model, from one pulsation cycle ($E_{\rm kin, max}^{i}$) to other ($E_{\rm kin, max}^{i+1}$). The corresponding kinetic energy growth rate, which we define as $\gamma=2(E_{\rm kin, max}^{i+1}-E_{\rm kin, max}^{i})/(E_{\rm kin, max}^{i+1}+E_{\rm kin, max}^{i})$, summed over $k$ pulsation cycles, $\gamma_k$, should be close to zero for the period-$k$ limit cycle. The method has thus additional advantage of pointing whether variations converge to limit cycle, or not. In the bottom panel of Fig.~\ref{fig.gamma} we plot both $\gamma_3$ and $\gamma_6$ for the considered model. For a reference, in the top panel we plot $\gamma_1$ for a singly-periodic $6332.0$\thinspace K model located well beyond the chaotic domain. The mean value of $\gamma_1$ for this model is zero, as expected. The small ($\sigma=3\cdot 10^{-5}$) fluctuations around the mean result partly from the interpolation (the time step is not an integer part of the period) and partly from the numerical noise which is relatively high in our models with low eddy viscosity (see Section~\ref{sec.hydro}). Analysis of the bottom panel of Fig.~\ref{fig.gamma} clearly points that the $6467.5$\thinspace K model display period-6 cycle behavior rather than period-3. This is further supported with the Fourier spectrum (Fig.~\ref{fig.fs_6467.5}). After pre-whitening with the fundamental mode and its harmonics (top panel of Fig.~\ref{fig.fs_6467.5}) a very strong signal is present at $1/3f_0$ and its harmonics. After pre-whitening we clearly detect signal at $1/6f_0$ (and harmonics), which is however, three order of magnitude weaker (middle panel of Fig.~\ref{fig.fs_6467.5}). After the next pre-whitening, signal is still present, which indicates that period-6 cycle is not strictly periodic, but displays irregular variation with very small amplitude. We note that only a fraction of Kelvin away chaotic band extends. 

\begin{figure}
\centering
\resizebox{\hsize}{!}{\includegraphics{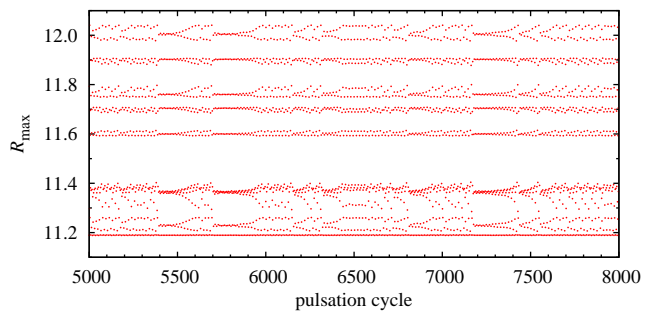}}
\caption{Values of maximum radii plotted over $3\,000$ pulsation cycles for $6365.0$\thinspace K model showing type-III intermittency.}
\label{fig.6365LOG}
\end{figure}

The other model ($6365.0$\thinspace K) is not strictly periodic either, which is clear already from the bifurcation diagram and first return map (Fig.~\ref{fig.rm_9}, brown points), which consists of nine separate and extended bands. Accordingly, strong signal at $1/9f_0$ (and harmonics) is present (top panel in Fig.~\ref{fig.fs_6365.0}). After pre-whitening, strong signal at $1/18f_0$ and its harmonics is present, which is not coherent however, as successive step of pre-whitening shows (bottom panel of Fig.~\ref{fig.fs_6365.0}). Fig.~\ref{fig.6365LOG}, showing the maximum radii over the $3\,000$ consecutive pulsation cycles, explains the origin of the $1/18f_0$ signal. The model switches intermittently between period-9 and period-18 cycle: starting from period-9 cycle the subharmonic cycle grows up to some amplitude and then model switches back to cycle-9 behaviour. This is type-III intermittency \citep[see][and Section~\ref{sec.logistic}]{PM80}, a result of collision between stable period-9 cycle and unstable  period-doubled (period-18) cycle arising from subcritical period doubling bifurcation. Type-III intermittency is rarely observed in dynamical systems and in most cases concerns switching between period-1 cycle and period-2 cycle \citep[e.g.][]{drb83}. The very similar behaviour as in the case of $6365.0$\thinspace K model was observed in electric circuits \citep[see][their fig.~21]{thamilmaran}. 

%%%%%%%%%%%%%%%%%%%%%%%%%%%%%%%%%%%%%%%%%%%%%%%%%%%%
\section{Largest Lyapunov exponents}\label{sec.lyap}
%%%%%%%%%%%%%%%%%%%%%%%%%%%%%%%%%%%%%%%%%%%%%%%%%%%%

One of the key features of chaotic system is its sensitivity to initial conditions, which can be quantified using Lyapunov exponents. For the chaotic attractor, the two initially nearby trajectories diverge at an exponential rate given by the largest Lyapunov exponent, $\lambda_1$. To determine $\lambda_1$ for our models we used the code and algorithm developed by \cite{RCL93} based on the original algorithm of \cite{GP83}. Below we briefly describe the underlying idea and refer the reader to original papers for details.

We first reconstruct the attractor dynamics using the method of delays. We use  the radius values equally spaced in time, $R_i$, to construct delay vectors, $\boldsymbol{x}_n$, in the embedding space:
\[\boldsymbol{x}_n=\big[R_{n-(m-1)J}, R_{n-(m-2)J},\ldots, R_{n} \big]. \]
$m$ is the embedding dimension and $J$ is time delay (or lag). The algorithm finds the close neighbours in the phase space, two points $\boldsymbol{x}_n$ and $\boldsymbol{x}_{n'}$ with sufficiently small distance between, $d_0=\boldsymbol{x}_n-\boldsymbol{x}_{n'}$. For chaotic system the distance should grow exponentially in time, so that $d_{0+l}=\boldsymbol{x}_{n+l}-\boldsymbol{x}_{n'+l}\approx d_0\exp(\lambda l)$, $\lambda$ corresponding to the largest Lyapunov exponent. Analysis of results for many pairs of close neighbours leads to a robust estimate of $\lambda_1$. 

For each model we have analysed only a section of the computed time series, of length $T$, expressed below as a multiple of the fundamental mode pulsation period. The default length of the analysed data is $\approx\!1\,000$ pulsation cycles. The embedding dimension cannot be smaller than the physical dimension of the attractor. Provided $m$ is high enough, its exact value should not affect the determined values of Lyapunov exponents. We find no systematic differences in the largest Lyapunov exponents computed assuming $m=4$ and $m=5$, and pick the latter value as the default (see also next paragraph). For the time delay we follow \cite{RCL93} and use $J$ for which autocorrelation function drops to $1-1/e$ of its initial value.

Before running computations for all models we checked the sensitivity of $\lambda_1$ to chosen values of $T$, $m$ and $J$, for one chaotic model of $6410.0$\thinspace K. For the chosen dense time sampling ($\delta t=0.01$d, more than $160$ points per pulsation cycle) the lag resulting from autocorrelation function is $J=27$. In the top row of Tab.~\ref{tab.lyap} we show $\lambda_1$ computed assuming default values of $T$, $m$ and $J$. Next, we have varied the values of these parameters and report the resulting largest Lyapunov exponents in the following rows of Tab.~\ref{tab.lyap}. We find that determination of $\lambda_1$ is robust and does not depend on exact values of the discussed parameters, provided that embedding dimension and lag are sufficiently high. Also, the chosen default length of the analysed time-series is sufficient, which we further verified computing $\lambda_1$ for all chaotic models, assuming the default values for $m$ and $J$, and two different lengths of time-series, the default one, of $1\,000$ pulsation cycles, and the shorter, of $670$ pulsation cycles. The results are plotted in the bottom part of the bifurcation diagram displayed in Fig.~\ref{fig.histo}, with thick black and thin blue lines respectively. Only for a few models a noticeable difference is present. We conclude that with $\sim 700-1\,000$ pulsation cycles the attractor dynamics is well probed.

\begin{table}
\caption{The largest Lyapunov exponents for the $6410.0$\thinspace K model computed with different values of time-series length, $T$, embedding dimension, $m$, and lag, $J$.}
\label{tab.lyap}
\centering
\begin{tabular}{rrrr}
$T$  & $m$ & $J$ & $\lambda_1$\thinspace [d$^{-1}$]\\
\hline
$1\,000$ &      $5$             &       $27$            & $0.217$\\
\hline
 $670$ & \multirow{2}{*}{$5$} & \multirow{2}{*}{$27$}   & $0.212$\\
$1\,400$ &                    &                         & $0.208$\\   
\hline
\multirow{4}{*}{$1\,000$} & $2$ & \multirow{4}{*}{$27$} & $0.136$\\
                      & $3$ &                           & $0.201$\\
                      & $4$ &                           & $0.212$\\
                      & $7$ &                           & $0.221$\\
\hline     
\multirow{4}{*}{$1\,000$} & \multirow{4}{*}{$5$} & $7$  & $0.184$\\
                      &                          & $17$ & $0.210$\\
                      &                          & $37$ & $0.220$\\
                      &                          & $47$ & $0.220$\\
\hline
\end{tabular}
\end{table}

The positive values of the largest Lyapunov exponents clearly establish the chaotic nature of our models.  The typical values of $\lambda_1$ (Fig.~\ref{fig.histo}) vary between $0.15$\thinspace d$^{-1}$ and $0.20$\thinspace d$^{-1}$. The largest values are slightly above $0.2$\thinspace d$^{-1}$ and are calculated for models between $6400.0$\thinspace K and $6420.0$\thinspace K. Within chaotic bands variation of $\lambda_1$ with effective temperature is not smooth. Large drops of $\lambda_1$ towards zero are seen, as expected, at the edges of periodic windows (for strictly periodic model $\lambda_1=0$). Smaller drops within chaotic domain may indicate a nearby periodic window, which was not detected because of too coarse resolution in effective temperature.

It is interesting to compare the computed values of $\lambda_1$ to those determined for two RV~Tau stars and one Mira-type variable. Results obtained for these stars are collected in Tab.~\ref{tab.lyap_comparison} together with $\lambda_1$ determination for one radiative chaotic model of \cite{kovb88}. Ranges of $\lambda_1$ rather than single values are given as determination of $\lambda_1$ from real stellar data is much more sensitive to the values of embedding dimension, lag, etc. The reader is referred to tables in original papers (references are given below Table) for details.

The values of largest Lyapunov exponent determined for our models are typically two orders of magnitude larger than values determined for stars or a radiative model listed in Tab.~\ref{tab.lyap_comparison}. On the other hand, the pulsation periods of our models (always between $1.5-1.7$\thinspace d) are order or two orders of magnitude shorter (see third column in Tab.~\ref{tab.lyap_comparison}). In the context of pulsating stars however, it is more appropriate to use the length of a single pulsation cycle (pulsation period) as a unit of time. When expressed in units of inverse pulsation cycles rather than d$^{-1}$, the values of largest Lyapunov exponents of our models are comparable to the values determined for RV~Tau/Mira-type stars. They are larger than for RV~Tau type stars, but a factor $\sim\!4$ smaller than for Mira type star. 

\begin{table}
\caption{Literature determinations of the largest Lyapunov exponents for RV~Tau/Mira-type stars and radiative model. Pulsation period is given in the third column (for RV~Tau stars undoubled period is given).}
\label{tab.lyap_comparison}
\centering
\begin{tabular}{lrrr}
star/type & $\lambda_1$\thinspace[$10^{-4}$\thinspace d$^{-1}$] & period [d] & Ref.\\
\hline
R~Cyg\thinspace/\thinspace Mira-type & $17-31$ & $\approx$430  & 1\\
R~Sct\thinspace/\thinspace RV~Tau    & $14-22$ & $\approx$70   & 2\\
AC~Her\thinspace/\thinspace RV~Tau   & $13-75$ & $\approx$37.5 & 3\\
D5200\thinspace/\thinspace model     & $34-55$ & 11.5 & 4\\
\hline
\end{tabular}
References: (1) \cite{ks03} (tab.~2), (2) \cite{bks96} (tab.~1), (3) \cite{kbsm98} (tab.~1), (4) \cite{skb96} (tab.~1).
\end{table}

%%%%%%%%%%%%%%%%%%%%%%%%%%%%%%%%%%%%%%%%%%%%%%%%%%%%%%
\section{Discussion}\label{sec.concl}
%%%%%%%%%%%%%%%%%%%%%%%%%%%%%%%%%%%%%%%%%%%%%%%%%%%%%%

The non-linear stellar pulsation equations we solve, form a much more complex system than classical chaotic systems discussed in the literature. Yet the resulting dynamical scenario is qualitatively similar to that arising from the iteration of the simplest logistic map (compare the bifurcation diagrams in Figs.~\ref{fig.log_bifurcation} and \ref{fig.histo} and animations attached with the online version of this paper as supporting information). Most of conclusions about the origin of dynamical phenomena found in our models are drawn based on the analogies between our models and simpler systems for which strict analytical reasoning is possible. However, for some of the phenomena we detect, we do not find a satisfactory analogy. The appearance of period-6 window, extending between $6459.0$\thinspace K and $6468.0$\thinspace K, is one of them. 

The scenario that is expected and that is encountered in many other systems, and is also present in the period-3 window (between $6421.0$\thinspace K and $6438.0$\thinspace K) is the following. First, a stable period-3 cycle is born together with unstable period-3 cycle. The stable branch undergoes a series of period doubling bifurcations to form three chaotic bands, which finally collide with the unstable period-3 cycle to form one chaotic band (Section~\ref{ssec.intercrises}). To the contrary, in the window at $\approx 6468.0$\thinspace K, a stable period-6 cycle, which looks like two stable period-3 cycles born very close to each other, emerges from the chaos. We are not aware of any bifurcation that may lead to such scenario and of any other system showing such behaviour. 

One of the possibilities is that in fact a pair of stable and unstable period-3 cycles is born, as expected, and stable cycle immediately undergoes a period doubling bifurcation. To check this, we have computed additional models in the interesting temperature range (with $0.1$\thinspace K-step in effective temperature), but they display either chaos or a period-6 behaviour (in Section~\ref{ssec.cs2} we analysed one of these models). If the proposed scenario indeed takes place it must occur in a temperature range narrower than $0.1$\thinspace K. 

The situation at $\approx 6468.0$\thinspace K looks even more complex. It seems that at this temperature we deal with discontinuity -- the bifurcation diagram (Fig.~\ref{fig.histo}) divides into two parts apparently decoupled from each other. On the cool side we clearly see a gradual evolution of the chaotic bands, divided, from time to time, by periodic windows. This gradual evolution seems to continue till $6468.0$\thinspace K, but not for the band extending at higher effective temperatures. This behaviour may result from the coexistence of two attractors in the system. Note that by default we initialized all the models along a sequence in the same manner (Section~\ref{sec.hydro}). It is possible that such initialization leads to different attractors for models cooler/hotter than $6468.0$\thinspace K. To check the possible existence of other stable attractor(s) (with different basins of attraction), we have repeated the computation for many models, but with several different initializations. In all cases however, we finally arrived at the same attractor as in the case of our default initialization. At the moment the cause and nature of bifurcation we observe at $6468.0$\thinspace K remains unclear.

The other phenomenon we have not discussed yet, is the appearance of chaos itself. The chaotic bands appear through a well understood period doubling route; the question is about the trigger. In the case of radiative models of \cite{bkov87} and \cite{kovb88}, analysis of the Floquet coefficients clearly shows that the first period doubling bifurcation is caused by the half-integer resonance between pulsation modes \citep[$5\!:\!2$ between fundamental and second overtone modes;][]{mb90}. The following cascade en route to chaos was not analysed. \cite{mb90} analysed a toy model of parametrically driven oscillator and showed that the first period doubling in such system results from the resonance and the following cascade is a result of increasing non-linearity. We note that the non-linearity may be the only cause of period doublings in classic chaotic systems void of internal resonances.

 We do not have appropriate tools (Floquet coefficients) to proof that half-integer resonance is responsible for the period doubling of single-periodic pulsation we observe at the cool and the hot sides of our computation domain. The closest resonances are (Fig.~\ref{fig.hr})  $\omega_{8}/\omega_{0}=9\!:\!2$ on the hot side  of the computation domain and, $\omega_{3}/\omega_{0}=5\!:\!2$ and $\omega_{5}/\omega_{0}=7\!:\!2$ on the cool side. The loci of the $\omega_{1}/\omega_{0}=3\!:\!2$ resonance, causing the period doubled pulsation detected in a single BL~Her star \citep{ssm12} is located more than 300\thinspace K-off the hot side of the computation domain considered here and likely plays no role. We cannot exclude the possibility that non-linearity is the only cause of the observed behaviours.

The appearance of periodic windows is also very interesting and important for stellar pulsation studies. As in the case of period-doubled pulsation, resonances were also invoked as a possible explanation. In a recent study \cite{pkm13} proposed that a $27\!:\!20$ resonance between the fundamental mode and the first overtone is responsible for a period-20 cycle behaviour they found in one of their RR~Lyrae models (their model H; for other model they propose a $14\!:\!19$ resonance). In this case however, a caution is needed, as inferences about the role of resonance are not based on firm theoretical grounds. The presence of first overtone cannot be deduced from the frequency spectrum. The role of resonance is most likely\footnote{\cite{pkm13} do not discuss in detail how the connection between the periodic pulsation and resonances is made.} deduced based on approximate coincidence of the model's location in the H-R diagram with the loci of the $27\!:\!20$ resonance determined with linear pulsation periods. Since pulsation periods change in the non-linear regime, and fine-tuning of such high-order resonance is difficult, claims on the possible role of resonances must be supported with other (dynamical) arguments. We are not aware of any studies showing that such high-order resonances may indeed have a noticeable effect on stellar pulsations. 

In a simpler explanation, period-$k$ behaviour detected in the models, is an intrinsic property of non-linear, chaotic system. The $\pm 1$\thinspace K neighbours of the discussed model of \cite{pkm13} show chaos and thus situation corresponds to periodic window within chaotic band, just as we report in this paper (Section~\ref{ssec.PW}), and as is found in many chaotic systems void of resonances. In chaotic systems the spectrum of periodic windows is dense \citep[it is one of the key properties of chaotic systems; e.g.][]{book.PJS}, but most of the windows are extremely narrow. Based on extreme similarity of our bifurcation diagram (Fig.~\ref{fig.histo}), to bifurcation diagrams for other systems, we conclude that also in the case of our computations the spectrum of periodic windows is most likely dense, but most of the windows are extremely narrow (in effective temperature). With the default $1$\thinspace K resolution of our model computation only few of the windows were detected (and most of them are narrower than $2$\thinspace K). Studying linear periods we find no tight connection between location of the periodic windows and the loci of high-order resonances between low order pulsation modes. We conclude that existence of periodic windows is an intrinsic property of non-linear system studied in this paper. There is no need to invoke resonances to explain them.

\section{Observability of the chaotic phenomena}\label{sec.obs}
%%%%%%%%%%%%%%%%%%%%%%%%%%%%%%%%%%%%%%%%%%%%%%%%%

Phenomena discussed in this paper are certainly exotic in the context of BL~Her stars. Except of period doubling phenomenon detected in one star \citep{igor11,ssm12}, BL~Her stars are single-periodic pulsators. We recall however, that existence of period-doubled BL~Her star was predicted $20$ years before its discovery \citep{bm92}. The future detection of other, more complex dynamical phenomena in these stars cannot be excluded. 

Period doubling and irregular brightness variations are common in higher luminosity siblings of BL~Her stars, namely in more luminous type II Cepheids of RV~Tau type and in semi-regular variables. Period doubling is a characteristic feature of RV~Tau stars which, in addition, show irregular variation. Chaos was reported in several stars of these types \citep{bks96,kbsm98,bkc04,ks03}. Our models suggest that other phenomena, intrinsic to chaotic dynamics, may also occur in these groups of stars. Particularly interesting would be the discovery of period-$k$ behaviour (with $k$ other than 2) and of intermittency. 

Period-$k$ behaviour may arise either within a period doubling cascade or within chaotic band. A possible period-4 behaviour in RV~Tau type star, supporting the period-doubling transition to chaos scenario, was reported by \cite{pollard}. Unfortunately, our unpublished analysis of the OGLE-III data for this star does not confirm the detection. Models indicate that in the case of period doubling cascade the domains of period-$k$ behaviour get narrower as $k$ increases, making the detection less probable for larger $k$. Also most of the periodic windows within chaotic domain are very narrow. Nevertheless, in our opinion, the firm detection of period-4 behaviour in RV~Tau stars is only a matter of time. 

From observers point of view, the strongest evidence for period-$k$ behaviour would be the presence of additional signal in the frequency spectrum, at $f_0/k$ and its harmonics. The difficulty arises for long-period variables, as very long time-series is necessary to get sufficient resolution in the sub-harmonic part of the frequency spectrum (i.e. in a range $[0,\,f_0]$). In addition, the effect may be obscured by irregular variability on top of period-$k$ behaviour, as is commonly observed in RV~Tau stars \citep[see eg. lightcurves in][]{igor11}. In these stars, a sporadic switching of the deep and shallow minima also occurs \citep{wallerstein} which is yet another difficulty in the analysis. Therefore, inspection of the time-series, folding the light curve with multiple of the basic period, are invaluable tools to search for the effect. This however requires not only long time-series, but also well sampled time-series. In this respect the projects aimed at long-term and regular monitoring of the sky or individual stars, such as OGLE \citep{ogle}, EROS \citep[e.g.][]{eros}, ASAS \citep{asas}, or programs led by amateur associations of variable star observers (e.g. AAVSO) are extremely important and should be continued as long as possible. It is worth noting that the observations of the only stars rigorously analysed for the presence of chaotic dynamic (i.e. those from Tab.~\ref{tab.lyap_comparison}) were all collected by amateur astronomers and in all cases covered more than 10 years (up to a century for R~Cyg).

Based on our rather restricted model survey we cannot predict the expected amplitude of brightness alternations within period-$k$ cycle. In period-doubled RV~Tau stars the amplitude of alternations vary from a hundredth of magnitude to few tenths of magnitude. More precise the photometry, larger the chance to detect the smaller effect.

Long-lasting and frequent observations of the stars are also crucial for the possible detection of  intermittency.  Models indicate that the intervals of apparently periodic/chaotic behaviour may last many tenths of pulsation cycles which, as pulsation periods of the luminous stars are long, requires a very long monitoring. Again, intermittency is expected in the narrow domains close to the edges of periodic windows and thus probability of its detection is certainly very small. 

\section{Summary}\label{sec.summary}

The BL~Her models discussed in this paper fall along a single stripe  of constant luminosity in the H-R diagram and cover a range of only $\approx\!150$\thinspace K. Yet they display a wealth of dynamical behaviours characteristic for deterministic chaos. Many of the discussed phenomena are detected for the first time in the context of stellar pulsation models. It was possible because our model survey was dedicated to study such phenomena -- a tiny step in effective temperature, sometimes as small as $0.1$\thinspace K (and $1$\thinspace K max), allowed to follow the dynamical evolution of the system from single-periodic pulsation, through period doubling cascade to well developed chaotic regime, and back to single-periodic pulsation. The chaotic regime turned out to be a gold-mine of interesting dynamical phenomena. We found several periodic windows (with cycle-3, 5, 6, 7 and 9 behaviours). We stress that the existence of periodic windows is not related to resonances among pulsation modes, but is an intrinsic property of a chaotic system. At the edges of the largest period-3 and period-6 windows we have found intermittent behaviour and crises bifurcations. Particularly interesting is intermittency -- a sporadic switching between two qualitatively different behaviours. In type-I intermittency intervals of apparently periodic (period-$k$, in general) behaviour are interrupted with bursts of chaos. In type-III intermittency the oscillations switch between two periodic cycles. In our models we detected switching between period-9 and period-18 cycles.

Detection of the discussed phenomena in the stars would be extremely interesting, however it requires a long and regularly sampled time series. The already available data from projects such as OGLE offer the best opportunity for a successful search.

\section*{Acknowledgments}
Model computations presented in this paper were conducted on the psk computer cluster in the Copernicus Centre, Warsaw, Poland. This research is supported by the Polish Ministry of Science and Higher Education through Iuventus+ grant (IP2012 036572) awarded to RS. PM is supported by the Polish National Science Centre through grant DEC-2012/05/B/ST9/03932.

\vspace{1cm}

\noindent{\bf SUPPORTING  INFORMATION}\\

\noindent Additional Supporting Information may be found in the online version of this article:\\

\noindent {\bf Animation 1:} The animation shows the evolution of first return map for logistic equation as control parameter $k$ is increased from $2.8$ to $4$. Note that smaller step is used within chaotic regime.\\

\noindent {\bf Animation 2:} The animation shows the evolution of first return map for BL~Her models discussed in the paper. Note the different, smaller effective temperature step at the edges of period-3 window.

%\bsp

\label{lastpage}

\end{document}